\documentclass{article}

\usepackage{amsfonts}
\usepackage{amsmath}
\usepackage{amssymb}
\usepackage{mathrsfs}

\title{The structure and interpretation of cosmology: Part II - The concept of creation in inflation and quantum cosmology}
\author{Gordon McCabe}

\def\eqalign#1{\,\vcenter{\openup.7ex\mathsurround=0pt
 \ialign{\strut\hfil$\displaystyle{##}$&$\displaystyle{{}##}$\hfil
 \crcr#1\crcr}}\,}

\begin{document}

\maketitle

\begin{abstract}

The purpose of the paper, of which this is part II, is to review,
clarify, and critically analyse modern mathematical cosmology. The
emphasis is upon mathematical objects and structures, rather than
numerical computations. Part II provides a critical analysis of
inflationary cosmology and quantum cosmology, with particular
attention to the claims made that these theories can explain the
creation of the universe.\hfill \break


\end{abstract}

\section{Introduction}

Part I of this paper (McCabe 2004) concentrated on general
relativistic cosmology, providing both critical analysis, and an
exposition of the mathematical structures employed, with the
purpose of demonstrating the great variety of possible universes
consistent with empirical data. Part II now provides a review and
critical analysis of inflation and quantum cosmology,
concentrating on the need to clarify concepts and, in particular,
to assess the claims made that inflation and quantum cosmology can
explain the creation\footnote{Whilst Grünbaum (1991) has suggested
substituting `origination' in the place of `creation', to avoid
conveying any theological connotations, the latter term is in such
widespread usage that it is employed in this paper, albeit without
the intent of conveying any of those connotations.} of the
universe.

\section{Inflation}

Inflationary cosmology postulates that the universe underwent a
period of acceleratory expansion in its early history due to the
existence of a scalar field $\phi$ with particular
characteristics. The scalar field is postulated to have an
equation of state $p = -\rho $,\footnote{An equation of state is a
functional expression which links the energy density $\rho$ of a
field with its pressure $p$.} and a particular type of potential
energy function $V(\phi)$. The inflationary scenarios postulate
that there was at least some patch of the early universe in which
this scalar field did not reside at the minimum of its potential
energy function, and in which the energy density of the scalar
field is dominated by its potential energy, $\rho = V(\phi)$.
Given the equation of state, this value of the scalar field
corresponds to a state of negative pressure, in which gravity is
effectively repulsive. A region of space in this so-called `false
vacuum' state undergoes exponential expansion until the scalar
field eventually falls into the minimum of its potential. After a
period of inflation, the false vacuum energy is converted into the
energy density of more conventional matter and radiation, and the
region of space which underwent inflation subsequently expands in
accordance with a Friedmann-Robertson-Walker (FRW) model.

In Guth's original 1981 proposal, the inflation was driven by a
scalar field which sat within a local, but not global, minimum of
its potential energy function. Whilst, in classical terms, this
state would be stable, quantum tunnelling would eventually cause
such a state to decay, thereby ending the inflationary expansion.
However, calculations indicated that this type of false vacuum
decay would cause density inhomogeneities inconsistent with
current observations. The \emph{new inflationary} scenario,
proposed both by Linde (1982), and the pairing of Albrecht and
Steinhardt (1982), solved this problem by proposing that the
scalar field which drives inflation sits atop a gentle plateau in
the potential energy function. With such a scalar field, inflation
takes place while the field slowly `rolls' into the global minimum
surrounding the plateau. This rolling process does not require
quantum tunnelling. Linde (1983a and 1983b) then proposed his
\emph{chaotic inflationary} scenario, in which the scalar field
potential can have a simple profile, with no plateau or local
minima, just a single global minimum at zero. In Linde's scenario,
inflation occurs because the field begins at a very high value,
and slowly `rolls' towards the global minimum.

Originally, the inflationary scalar field was identified as the
Higgs field from the Grand Unified Theories (GUTs) of particle
physics, and inflation was triggered by spontaneous symmetry
breaking of the GUT gauge symmetry. Grand Unified Theories
hypothesise that at energies of about $10^{14}$GeV, the
electroweak and strong forces merge into a single unified force.
Such theories also postulate the existence of Higgs fields. The
new inflationary scenario postulated that as the universe
approached the age of $10^{-35}s$, the matter in the universe was
in its Grand Unified phase, with the electroweak and strong forces
unified, and with the GUT Higgs fields all set to zero. As the
universe expanded, it cooled, and after $10^{-35}s$ the
temperature of the universe dropped below the level at which the
electroweak and strong forces are unified. This sudden change in
the state of the matter in the universe is called the GUT `phase
transition'. If such a phase transition occurred rapidly when the
temperature fell to the critical value, there would be no
inconsistency with FRW cosmology, (Blau and Guth 1987, p528).
However, the new inflationary scenario proposed that the universe
underwent supercooling at the GUT phase transition. In other
words, it was proposed that the phase transition occurred slowly
compared with the rate of cooling. As a result of supercooling, it
was hypothesised that the energy density of the universe became
dominated by the energy density of the GUT Higgs fields, and the
thermal component of the energy density became negligible in
comparison. A region of space in such a false vacuum state would
undergo exponential expansion until the Higgs fields fall into a
`true vacuum state'. In terms of the classical theory, the set of
true vacuum states is simply defined by the global minimum of the
potential energy function.

Practitioners of inflation now assert that it is not possible to
identify the scalar field responsible for inflation with the Higgs
field of GUTs, ``since the potential of such a scalar field is too
steep," (Brandenberger 2002, p4). Inflation driven by the
potentials of GUT Higgs fields purportedly results in excessive
density perturbations; the consequent amplitude of the
anisotropies in the cosmic microwave background radiation exceed
that which is actually observed. The scalar field responsible for
inflation is now widely referred to as the `inflaton', and is
often considered in abstraction from particle physics.

The period of acceleratory expansion postulated in inflation has
the consequence that the presently observable universe came from a
region sufficiently small that it would have been able to reach
homogeneity and thermal equilibrium by means of causal processes
before the onset of inflation. Inflation thereby solves the
so-called `horizon problem' of FRW cosmology. The apparent
homogeneity of our observable universe has to be built-in to the
initial conditions of a FRW model, dictating the choice of a
locally isotropic and locally homogeneous $3$-dimensional
Riemannian manifold to represent the spatial universe. Regions of
space on opposite sides of our observable universe have the same
average temperature and density, even though, in a FRW model, they
have always lain beyond each other's particle horizons. Under
inflation, the observable universe comes from a region which would
have been able to reach a homogeneous state despite starting from
a possibly heterogeneous initial state.

However, to regard the horizon phenomenon in the FRW models as a
`problem' betrays a methodological assumption that one can only
explain things with causal processes rather than by initial
conditions. The universe could, quite simply, have been
homogeneous from the outset.

\hfill \break

A defining characteristic of inflation is that the energy density
of the inflating region is constant during the period of
acceleratory expansion. The energy density is maintained at the
false vacuum energy density, $\rho_f$, throughout the period of
inflation.\footnote{This characteristic plays a key role in the
ideas for universe creation `in a laboratory'.} Although the
calculated energy density at the onset of inflation was huge, at,
say, $\rho_f \approx 10^{73} g \, cm^{-3}$, by integrating it over
the very small region which became the observable universe, one
gets a relatively small total energy. Thus, Guth and Steinhardt
assert that ``probably the most revolutionary aspect of the
inflationary model is the notion that all the matter and energy in
the observable universe may have emerged from almost nothing,"
(1989, p54).

One begins with a region of very small volume at the time
inflation was triggered. During inflation, the scale factor of
this region increases enormously, but the energy density remains
constant. The huge increase in the scale factor means a huge
increase in the volume of the region. Thus, during inflation, the
total (non-gravitational energy) of the region increases. At the
completion of inflation, the energy density is the same that it
was to begin with, but the region has a much greater volume.
Integrating the energy density over a much larger domain, one gets
a larger total energy.

The consequence of this, as Guth and Steinhardt explain, is that
``essentially all the non-gravitational energy of the [observable]
universe is created as the false vacuum undergoes its accelerated
expansion. This energy is released when the phase transition takes
place, and it eventually evolves to become everything that we see,
including the stars, the planets, and even ourselves," (1989,
p54).

It is important to emphasise that inflation only entails the
\emph{observable} universe to have been created from a very small
initial amount of energy. Inflation does not entail that the
entire universe was created from almost nothing. The entire
spatial universe could have either compact or non-compact
topology, and could therefore be either of finite volume, or of
infinite volume. In contrast, the observable spatial universe is
definitely of finite volume. This entails that the total amount of
non-gravitational energy within the observable spatial universe
must be finite. If the entire spatial universe is compact, the
total amount of non-gravitational energy in the spatial universe
will also be finite, but if the entire spatial universe is
non-compact, the total non-gravitational energy in the universe
could be infinite. It is only if the entire spatial universe is
compact, and therefore of finite volume, like the observable
universe, that the entire universe could have been created from
`almost nothing'.

Guth and Steinhardt conclude that ``the inflationary model offers
what is apparently the first plausible scientific explanation for
the creation of essentially all the matter and energy in the
observable Universe," (1989, p54). They acknowledge that ``it is
then tempting to go one step further and speculate that the entire
universe evolved from literally nothing. The recent developments
in cosmology strongly suggest that the universe may be the
ultimate free lunch," (1989, p54).

This, of course, is where quantum cosmology enters. Blau and Guth
claim that in the scenarios proposed by Vilenkin and Linde, ``the
universe tunnels directly from a state of `absolute nothingness'
into the false vacuum," and that Hartle and Hawking ``have
proposed a unique wave function for the universe, incorporating
dynamics which leads to an inflationary era," (1987, p556). These
latter claims are over-optimistic, and are typical of the way in
which quantum cosmology is often invoked as a \textit{deus ex
machina} to explain the initial conditions which are necessary for
inflation to occur.

Despite this criticism, one can endorse the interpretation of Guth
and Steinhardt, that inflation is able to explain how almost all
the non-gravitational energy in our observable universe was
created. Inflation, however, clearly cannot explain how space and
time were created, and it cannot explain how the initial seed of
energy was created. Inflation cannot produce physical something
from physical nothing. Inflation could, quite conceivably, be a
vital cog in a universe creation theory, but it cannot on its own
explain why there is physical something rather than physical
nothing.

\hfill \break

The hypothetical false vacuum state which drives inflation is
distinct from the true vacuum, which, in classical terms, is
defined by the global minimum of the potential energy function.
Whilst it is conventional to set the global minimum in the
classical theory to zero, according to quantum theory the true
vacuum state of a field does not have zero energy. In the quantum
vacuum, it is believed that virtual particle-antiparticle pairs
are constantly created and annihilated. It is believed that the
virtual pairs are created \textit{ex nihilo}, and physicists speak
of the quantum `fluctuations' of the vacuum.

The nature of the quantum vacuum has inspired a number of universe
creation scenarios. For example, in 1973 Edward P. Tryon proposed
that our universe was created as a spontaneous quantum fluctuation
of some pre-existing `vacuum'. Tryon conjectured that all
conserved quantities have a net value of zero for the universe as
a whole. Noting that in Newtonian theory, the gravitational
potential energy is negative, he proposed that there might be a
sense in which the negative gravitational energy of the universe
cancels the positive mass-energy. He calculated that this
might/would be the case if the average density of matter matches
the critical density (1973, p396), although he also seemed to
predict that the universe is closed (1973, p397).

Tryon's idea still finds favour today. Guth suggests that the
energy created during inflation ``comes from the gravitational
field. The universe did not begin with this energy stored in the
gravitational field, but rather the gravitational field can supply
the energy because its energy can become negative without bound.
As more and more positive energy materializes in the form of an
ever-growing region filled with a high-energy scalar field, more
and more negative energy materializes in the form of an expanding
region filled with a gravitational field. The total energy remains
constant at some very small value, and could in fact be exactly
zero," (Guth 2004, p5-6). However, Tryon's idea runs aground on a
fact Guth mentions in a footnote: ``In general relativity there is
no coordinate-invariant way of expressing the [gravitational]
energy in a space that is not asymptotically flat, so many experts
prefer to say that the total energy is undefined," (ibid., p6). As
Wald points out, ``it has long been recognized that there is no
meaningful local notion of gravitational energy density in general
relativity," (Wald 2001, p20).

In 1978, Brout \textit{et al} adopted the idea of an initial
microscopic quantum fluctuation, but added the idea that the
initial state of matter was one with a large negative pressure,
which resulted in exponential expansion of the initial fluctuation
into an open universe. The creation of an open universe featured
in a paper by J.R.Gott in 1982, and in the same year, the papers
of Atkatz-Pagels, and Vilenkin addressed the creation of closed
universes. Subsequently, Tryon argued (1992) that inflation can be
combined with the notion of a quantum vacuum fluctuation to
explain the creation of a universe.

On the one hand, Tryon believes that the notion of a quantum
fluctuation alone is sufficient to explain the creation of our
universe, stating that ``although quantum fluctuations are
typically microscopic in scale, no principle limits their
potential size and duration, provided that conservation laws are
respected. Hence, given sufficient time, it seems inevitable that
a universe with the size and duration of ours would spontaneously
appear as a quantum fluctuation," (Tryon, 1992, p571). On the
other hand, he acknowledges that ``large (and long-lived)
universes intuitively seem much less likely than smaller ones," so
he then suggests that inflation could transform an initial
microscopic fluctuation into a large universe. He asserts that
``inflation greatly enhances the plausibility of creation ex
nihilo," and concludes that ``quantum uncertainties suggest the
instability of nothingness...inflation might have converted a
spontaneous, microscopic quantum fluctuation into our Cosmos,"
(1992, p571).

Tryon fails to establish a clear distinction between the possible
creation of the material universe from a pre-existing `empty'
space-time, and the possible creation of space, time, and matter
from physical nothing, the empty set $\emptyset$. In 1973, Tryon
imagined our universe as ``a fluctuation of the vacuum, the vacuum
of some larger space in which our Universe is imbedded," (Tryon,
1973, p397). This statement seems to indicate that Tryon was
thinking of creation from a pre-existing, empty space-time. It
seems to indicate that the `vacuum' Tryon refers to is the matter
field vacuum of a pre-existing empty space-time. Subsequently,
Tryon stated his proposal more carefully, asserting that ``the
universe was created from nothing as a spontaneous quantum
fluctuation of some pre-existing vacuum or state of nothingness,"
(1992, p570). From the latter statement, it seems that Tryon now
contemplates creation from either a pre-existing empty space-time,
or from literally nothing, the empty set.

Even then, however, Tryon argues that ``given sufficient time"
(1992, p571) quantum fluctuations will yield a universe. This
echoes the 1973 proposal that our universe ``is simply one of
those things which happen from time to time." Both comments
indicate that Tryon considers time to exist before the
hypothetical creation of our universe as a vacuum fluctuation.
This is consistent with the idea that a universe is created as a
quantum fluctuation in a pre-existing space-time. It is
inconsistent with the idea that a universe is created from
physical nothing, the empty set.

Whilst inflation on its own could only explain the existence of
\emph{almost all} the matter and non-gravitational energy in our
universe, by combining inflation with the idea of a quantum
fluctuation in a pre-existing space-time, one might be able to
explain the existence of \emph{all} the matter and
non-gravitational energy in our universe. One might suggest that
there existed an initial space-time in which the matter fields
were in their true vacuum states. One might then imagine that some
fluctuation of this quantum vacuum created a small region of space
in which the inflaton scalar field possesses the necessary initial
state for inflation to ensue. The small initial quantum
fluctuation would be transformed into a fully-fledged universe.
Inflation would transform the small initial amount of
non-gravitational energy into enough matter and non-gravitational
energy for a universe larger than our own observable universe.

It is important to note that two distinct types of vacuum are at
work in such a scenario. Quantum fluctuations of the true vacuum
would create a small amount of non-gravitational energy, `almost
nothing', and then the properties of the false vacuum would
create, from `almost nothing', sufficient non-gravitational energy
for a universe replete with galaxies.

Obviously, such a scenario would only explain the creation of the
material universe. It would not be creation from physical nothing,
because there would be a pre-existing space-time. Tryon's idea
would not incorporate inflation into a theory which explained why
there is physical something rather than physical nothing. Tryon's
idea would, at best, incorporate inflation into a theory which
explained why there is some matter and energy, rather than no
matter and energy.

Given the current notion of physical space and the current notion
of the quantum vacuum, the existence of the quantum vacuum is
contingent. It is not a contradiction to imagine the existence of
space and the non-existence of the quantum vacuum. It is false to
claim that truly empty space, with zero energy, is impossible.
There is nothing in the current notion of physical space that
entails the presence of the quantum vacuum. So long as space is
represented by a differential manifold, and mass-energy is
represented by fields on a manifold, it will be possible to
imagine empty space. It may well be true that there is no
operational procedure which can make a region of space completely
empty, but this does not mean that it is impossible for space to
be empty. It might also be operationally impossible to change the
dimension of physical space, but that does not mean that it is
impossible for physical space to be other than $3$-dimensional.

Some theory in the future may represent the universe in a way that
makes space-time and mass-energy conceptually inseparable, and it
may then follow from the nature of space-time that the quantum
vacuum exists. However, if this were to be the case, there would
no longer be the twofold question of how a material universe could
have been created from empty space, and how empty space could have
been created from physical nothing. One would have the single
question of how the physical universe could have been created from
physical nothing. Hence, the notion of the quantum vacuum cannot
entail the existence of the material universe. If space-time and
mass-energy are conceptually separable, then the presence of the
quantum vacuum is merely contingent, hence it cannot entail that
empty space must create a material universe. Alternatively, if
space-time and mass-energy are conceptually inseparable, then an
explanation for the existence of the material universe requires an
explanation of how the material universe was created from physical
nothing, and the quantum vacuum cannot achieve this.

\section{Quantum cosmology}

In canonical general relativity, expressed in terms of the
`traditional' variables, a configuration of the spatial universe
is given by a $3$-dimensional manifold $\Sigma$, equipped with a
Riemannian metric tensor field $\gamma$, and a matter field
configuration $\phi$. The full configuration space of general
relativity would be the set of all possible pairs $(\gamma,\phi)$
on all possible $3$-manifolds $\Sigma$.\footnote{In terms of
Ashtekar's `new variables', the geometrical configuration space is
not the space of metrics on $\Sigma$, but the space of connections
upon an $SU(2)$-principal fibre bundle over $\Sigma$, (Baez
1995).} In canonical quantum gravity, the main object of interest
is a state vector $\Psi$, a functional upon the configuration
space which satisfies the Wheeler-DeWitt equation.

In path-integral quantum gravity, expressed in terms of the
traditional variables, the main object of interest is a transition
from an initial configuration $(\Sigma_i,\gamma_i,\phi_i)$ to a
final configuration $(\Sigma_f,\gamma_f,\phi_f)$. The interest
lies in defining and calculating a propagator
$K(\Sigma_i,\gamma_i,\phi_i;\Sigma_f,\gamma_f,\phi_f)$. In
contrast with path-integral non-relativistic quantum mechanics,
there are no overt time labels associated with either the initial
or final configuration.

To calculate the propagator
$K(\Sigma_i,\gamma_i,\phi_i;\Sigma_f,\gamma_f,\phi_f)$, one might
expect to introduce the set
$\mathscr{P}_L(\Sigma_i,\gamma_i,\phi_i
;\Sigma_f,\gamma_f,\phi_f)$, of all $4$-dimensional Lorentzian
space-times which interpolate between $(\Sigma_i,\gamma_i,\phi_i)$
and $(\Sigma_f,\gamma_f,\phi_f)$. Whilst classical general
relativity requires that a space-time satisfy the classical
dynamical equations, the Einstein field equations, quantum gravity
introduces the set of all kinematically possible interpolating
space-times, irrespective of whether they satisfy the Einstein
field equations.

Each interpolating space-time history is a 4-dimensional
Lorentzian manifold-with-boundary $(\mathcal{M},g)$. The boundary
of each $\mathcal{M}$ must consist of the disjoint union of
$\Sigma_i$ and $\Sigma_f$. In addition, the restriction of the
Lorentzian metric $g$ to the boundary components must be such that
$g| \Sigma_i = \gamma_i$ and $g | \Sigma_f = \gamma_f$. Each
interpolating space-time must be equipped with a smooth matter
field history $\Phi$, which satisfies the conditions $\Phi |
\Sigma_i = \phi_i$ and $\Phi | \Sigma_f = \phi_f$.

The initial $3$-manifold $\Sigma_i$ need not be homeomorphic with
the final $3$-manifold $\Sigma_f$. Hence, the transition from an
initial configuration $(\Sigma_i,\gamma_i,\phi_i)$ to a final
configuration $(\Sigma_f,\gamma_f,\phi_f)$ could be a topology
changing transition.

The notion of topology change is closely linked with the concept
of cobordism. When a pair of $n$-manifolds, $\Sigma_1$ and
$\Sigma_2$, constitute disjoint boundary components of an $n+1$
dimensional manifold, $\Sigma_1$ and $\Sigma_2$ are said to be
cobordant. It is a valuable fact for path-integral quantum gravity
that any pair of compact $3$-manifolds are cobordant, (Lickorish
1963). Not only that, but any pair of compact Riemannian
$3$-manifolds, $(\Sigma_1,\gamma_1)$ and $(\Sigma_2,\gamma_2)$,
are `Lorentz cobordant', (Reinhart 1963). i.e. There exists a
compact $4$-dimensional Lorentzian manifold $(\mathcal{M},g)$,
with a boundary $\partial \mathcal{M}$ which is the disjoint union
of $\Sigma_1$ and $\Sigma_2$, and with a Lorentzian metric $g$
that induces $\gamma_1$ on $\Sigma_1$, and $\gamma_2$ on
$\Sigma_2$.

This cobordism result is vital because it confirms that topology
change is possible. Even when $(\Sigma_1,\gamma_1)$ and
$(\Sigma_2,\gamma_2)$ are compact Riemannian $3$-manifolds with
different topologies, there exists an interpolating space-time.

With each kinematically possible interpolating history, one can
associate a real number, the action $A$

$$
A = \frac{1}{16}\pi G \int_\mathcal{M}S \sqrt{-g}\,d^4 x +
\frac{1}{8} \pi G \int_{\partial \mathcal{M}}Tr K
\sqrt{\gamma}\,d^3 x + C +\int_\mathcal{M}L_m \sqrt{-g}\,d^4 x \,.
$$ $S$ is the scalar curvature, $K$ is the extrinsic curvature
tensor, and $L_m$ is the matter field Lagrangian density.

One `weights' each possible space-time history with a unimodular
complex number $exp(iA/\hbar)$. Whilst $A: \mathscr{P}_L
\rightarrow \mathbb{R}^1$ is an unbounded function on the
path-space $\mathscr{P}_L$, the mapping $exp(iA/\hbar):
\mathscr{P}_L \rightarrow S^1 \subset \mathbb{C}^1$ is a bounded
function.

One could then define the propagator of quantum gravity as:

$$
K(\Sigma_i,\gamma_i,\phi_i;\Sigma_f,\gamma_f,\phi_f) =
\int_{\mathscr{P}_L} exp(iA/\hbar) d\mu \,.
$$

It has been claimed that in quantum gravity, the creation of a
universe from nothing would simply correspond to the special case
where $(\Sigma_i,\gamma_i,\phi_i) = \emptyset$. If this were so,
then the probability amplitude or probability of a transition from
nothing to a spatial configuration $(\Sigma_f,\gamma_f,\phi_f)$
would be given by

$$
K(\emptyset;\Sigma_f,\gamma_f,\phi) = \int_{\mathscr{P}_L}
exp(iA/\hbar) d\mu \, ,
$$ where $\mathscr{P}_L$ is an abbreviation here for
$\mathscr{P}_L(\emptyset;\Sigma_f,\gamma_f,\phi_f)$, the set of
all Lorentzian $4$-manifolds $(\mathcal{M},g)$ and matter field
histories $\Phi$ which have a single boundary component $\partial
\mathcal{M} = \Sigma_f$ on which $g$ induces $\gamma_f$, and
$\Phi$ induces $\phi_f$.

Unfortunately, there are serious technical problems with the
definition of the propagator by a Lorentzian path-integral.
Firstly, if one permits $\mathscr{P}_L$ to include non-compact
space-times, then the action integral can diverge for some of
these space-times. For example, if a non-compact space-time is
homogeneous, then the action integral diverges. Because a
non-compact homogeneous space-time has no well-defined action $A$,
it cannot be assigned a weight $exp(iA/\hbar)$. An asymptotically
flat space-time is a notable case of a non-compact space-time for
which the action integral is finite, but asymptotically flat
geometry is a special case, and is of no cosmological relevance.

Secondly, $\mathscr{P}_L$ is not finite-dimensional, and no
satisfactory measure has been found on $\mathscr{P}_L$. In the
absence of a satisfactory measure on $\mathscr{P}_L$, integration
over $\mathscr{P}_L$ is not well-defined. Although the integrand
$exp(iA/\hbar)$ is a bounded function, when it is expanded into
its real-imaginary form, $exp(iA/\hbar) = \cos A/\hbar + i \sin
A/\hbar$, it is clearly oscillatory. Thus, even if one attempted
to approximate the propagator by an integral over a
finite-dimensional subset of $\mathscr{P}_L$, the integral would
not be finite unless one integrated over a compact subset of
$\mathscr{P}_L$. One attempt to avoid these difficulties is the
so-called `Euclidean' path-integral approach to quantum gravity.
In this approach, the propagator
$K(\Sigma_i,\gamma_i,\phi_i;\Sigma_f,\gamma_f,\phi_f)$ is defined
to be an integral over
$\mathscr{P}_R(\Sigma_i,\gamma_i,\phi_i;\Sigma_f,\gamma_f,\phi_f)$,
the set of all compact Riemannian $4$-manifolds and matter field
histories which interpolate between $(\Sigma_i,\gamma_i,\phi_i)$
and $(\Sigma_f,\gamma_f,\phi_f)$. It would clearly be more
appropriate to refer to this approach as the Riemannian
path-integral approach to quantum gravity.

A `Euclidean' action $A_E$ is associated with each interpolating
history, and one assigns a weight of $exp(-A_E/\hbar)$ to each
such interpolating history. The propagator is then defined to be

$$
K(\Sigma_i,\gamma_i,\phi_i;\Sigma_f,\gamma_f,\phi_f) =
\int_{\mathscr{P}_R} exp(-A_E/\hbar) d\mu \,.
$$

A Riemannian manifold $(\mathcal{M},g)$ has the helpful property
that if either i) $\mathcal{M}$ is compact, or ii)
$(\mathcal{M},g)$ is homogeneous, then $(\mathcal{M},g)$ must be
geodesically complete. Hence, by integrating over compact
Riemannian 4-geometries, one would be integrating over
geodesically complete geometries; one would be integrating over
`non-singular' 4-geometries. For this reason, advocates of
Euclidean path-integrals tend to claim that their approach avoids
the singularities of classical cosmology. Note, however, that
quantum cosmology replaces an individual space-time manifold with
objects such as wave-functions and propagators, so the issue of a
singularity in the geometry is no longer so pertinent.

If the creation of a universe from nothing corresponds to the
special case in which $(\Sigma_i,\gamma_i,\phi_i) = \emptyset$,
then in the Euclidean approach the probability amplitude or
probability of a transition from nothing to a spatial
configuration $(\Sigma_f,\gamma_f,\phi_f)$ would be given by
integrating only over the compact Riemannian $4$-manifolds and
matter field histories $\mathscr{P}_R =
\mathscr{P}_R(\emptyset;\Sigma_f,\gamma_f,\phi_f)$:

$$
K(\emptyset;\Sigma_f,\gamma_f,\phi_f) = \int_{\mathscr{P}_R}
exp(-A_E/\hbar) d\mu \,.
$$

Unfortunately, the Euclidean action $A_E$ is not positive
definite; $A_E$ can be negative. Moreover, there is no lower bound
on the value that the Euclidean action can take. Thus, the
integrand in the path integral, $exp(-A_E/\hbar) =
1/exp(A_E/\hbar)$ can `blow up exponentially'. This means that the
integrand in a Riemannian path-integral can be an unbounded
function. If one attempted to approximate the propagator by
integrating $exp(-A_E/\hbar)$ over a finite-dimensional subset of
$\mathscr{P}_R$, then the integral would not be finite unless one
used a special measure. In the Euclidean approach it has been
suggested that the transition amplitudes
$K(\emptyset;\Sigma,\gamma,\phi)$ can be approximated by summation
over compact Riemannian $4$-geometries which are saddle points of
the action $A_E$. However, even if there is a way to approximately
calculate the transition amplitudes of quantum gravity, it is
highly debatable whether the transition amplitudes
$K(\emptyset;\Sigma,\gamma,\phi)$ could be interpreted as creation
\textit{ex nihilo} amplitudes.

In the case of the Lorentzian approach, the first problem is that
compact Lorentzian space-times with only a single compact boundary
component, are time non-orientable. This means that the single
compact boundary cannot be treated as a final boundary, at which
the region of space-time ends. It is equally legitimate to treat
it as a boundary at which the region of space-time begins.

Suppose instead that one uses a collection of non-compact,
time-orientable Lorentzian space-times which end at
$(\Sigma,\gamma,\phi)$, and which have no past boundary. Each one
of these space-times `creates' $(\Sigma,\gamma,\phi)$ from a prior
region of space-time. Thus, all the space-times which determine
the purported creation \textit{ex nihilo} probability of
$(\Sigma,\gamma,\phi)$, `create' $(\Sigma,\gamma,\phi)$ from a
prior region of space-time; they do not individually create
$(\Sigma,\gamma,\phi)$ from nothing $\emptyset$. Indeed, some
space-times which terminate with $(\Sigma,\gamma,\phi)$ are
past-infinite. Thus, space-times which exist for an infinite time
before reaching $(\Sigma,\gamma,\phi)$ would contribute to the
probability of creating $(\Sigma,\gamma,\phi)$ from nothing!

Similarly, in the Euclidean approach, all the Riemannian
$4$-geometries which determine the purported creation \textit{ex
nihilo} probability of $(\Sigma,\gamma,\phi)$, `create'
$(\Sigma,\gamma,\phi)$ from a region of 4-dimensional space; they
do not individually create $(\Sigma,\gamma,\phi)$ from nothing
$\emptyset$.

These are strong reasons to doubt that
$K(\emptyset;\Sigma,\gamma,\phi)$ could be interpretable as a
creation \textit{ex nihilo} probability amplitude in either the
Lorentzian or the Euclidean approach. In the Lorentzian approach,
when speaking of space-times with no past boundary, it is
syntactically acceptable to say that the past boundary component
is empty, $\emptyset$, but one should not think of $\emptyset$ as
a special type of past boundary; it is no past boundary at all. In
the Euclidean approach, when speaking of $4$-dimensional spaces
with no second boundary component, it is syntactically acceptable
to say that the second boundary component is empty, $\emptyset$,
but again one should not think of $\emptyset$ as a special type of
second boundary; rather, it is no second boundary at all. A
boundary of a manifold must be a topological space, and amongst
other things, a topological space must be a non-empty set.
$\emptyset$ is the empty set, hence $\emptyset$ cannot be a
topological space, which entails that $\emptyset$ cannot be the
boundary of a manifold. Cobordism is an equivalence relation
between manifolds, hence it is not possible for any manifold to be
cobordant with the empty set $\emptyset$.

Space-times which have no past boundary are not space-times which
begin with the empty set $\emptyset$. As Grünbaum complains,
``What...is temporally `initial' about an empty set...?
Apparently, the empty set in question is verbally labelled to be
`initial' by mere definitional fiat," (Grünbaum 1991, Section C).
An integration or summation over space-times with no past
boundary, can only be interpreted as the probability of
$(\Sigma_f,\gamma_f,\phi_f)$ arising from anything, not the
probability of $(\Sigma_f,\gamma_f,\phi_f)$ arising from nothing.
The absence of a past boundary merely signals the absence of a
restriction upon the ways in which $(\Sigma_f,\gamma_f,\phi_f)$
can come about. Every space-time in
$\mathscr{P}_L(\Sigma_i,\gamma_i,\phi_i;\Sigma_f,\gamma_f,\phi_f)$,
for each $(\Sigma_i,\gamma_i,\phi_i)$, is a subset of at least one
space-time in $\mathscr{P}_L(\emptyset;\Sigma_f,\gamma_f,\phi_f)$.
Every space-time in
$\mathscr{P}_L(\Sigma_i,\gamma_i,\phi_i;\Sigma_f,\gamma_f,\phi_f)$
is part of at least one space-time in
$\mathscr{P}_L(\emptyset;\Sigma_f,\gamma_f,\phi_f)$ which extends
further into the past, beyond $(\Sigma_i,\gamma_i,\phi_i)$. It is
in this sense that the absence of a past boundary merely signals
the absence of a restriction upon the ways in which
$(\Sigma_f,\gamma_f,\phi_f)$ can come about. The set of Lorentzian
space-times $\mathscr{P}_L(\emptyset;\Sigma_f,\gamma_f,\phi_f)$
contains all the possible past histories that lead up to
$(\Sigma_f,\gamma_f,\phi_f)$, whereas
$\mathscr{P}_L(\Sigma_i,\gamma_i,\phi_i;\Sigma_f,\gamma_f,\phi_f)$
contains the past histories which are truncated at the spatial
configuration $(\Sigma_i,\gamma_i,\phi_i)$. An integration or
summation over space-times with no past boundary cannot be
interpreted as the probability of a transition from $\emptyset$ to
$(\Sigma_f,\gamma_f,\phi_f)$.

Similarly, in the Euclidean approach, an integration or summation
over $4$-dimensional spaces in which $(\Sigma_f,\gamma_f,\phi_f)$
is the only boundary component, cannot be interpreted as the
probability of a transition from $\emptyset$ to
$(\Sigma_f,\gamma_f,\phi_f)$. The absence of another boundary
component merely signals the absence of a restriction upon the
$4$-dimensional spaces which possess $(\Sigma_f,\gamma_f,\phi_f)$
as a boundary. Every Riemannian $4$-geometry in
$\mathscr{P}_R(\Sigma_i,\gamma_i,\phi_i ;
\Sigma_f,\gamma_f,\phi_f)$, for each $(\Sigma_i,\gamma_i,\phi_i)$,
is a subset of at least one Riemannian $4$-geometry in
$\mathscr{P}_R(\emptyset;\Sigma_f,\gamma_f,\phi_f)$. Every
Riemannian $4$-geometry in $\mathscr{P}_R(\Sigma_i,\gamma_i,\phi_i
; \Sigma_f,\gamma_f,\phi_f)$ is part of at least one Riemannian
$4$-geometry in
$\mathscr{P}_R(\emptyset;\Sigma_f,\gamma_f,\phi_f)$ which extends
to a greater volume, beyond $(\Sigma_i,\gamma_i,\phi_i)$. The set
of Riemannian 4-geometries
$\mathscr{P}_R(\emptyset;\Sigma_f,\gamma_f,\phi_f)$ contains all
the possible 4-dimensional spaces which possess
$(\Sigma_f,\gamma_f,\phi_f)$ as a boundary, whereas
$\mathscr{P}_R(\Sigma_i,\gamma_i,\phi_i ;
\Sigma_f,\gamma_f,\phi_f)$ contains all those that are truncated
at the spatial configuration $(\Sigma_i,\gamma_i,\phi_i)$.

To counter these arguments, one could argue that the space-times
or $4$-dimensional spaces being used would only play a part in the
theoretical calculation of the transition probabilities, and would
not play a part in any actual physical process. One could argue
that the transition from $\emptyset$ to some
$(\Sigma,\gamma,\phi)$ only takes place at the quantum level, not
at the level of the individual classical space-times or
$4$-dimensional spaces which are used to calculate the probability
of the quantum event. One could argue that the only thing which
happens physically is a transition from $\emptyset$ to some
$(\Sigma,\gamma,\phi)$. The fact that the space-times used in the
Lorentzian approach cannot be said to begin with the empty set,
and the fact that they individually create $(\Sigma,\gamma,\phi)$
from a prior region of space-time, does not entail that they
cannot be used to calculate the probability of a transition from
$\emptyset$ to $(\Sigma,\gamma,\phi)$. The fact that the
$4$-dimensional spaces used in the Euclidean approach cannot be
said to individually create $(\Sigma,\gamma,\phi)$ from the empty
set, and the fact that they individually `create'
$(\Sigma,\gamma,\phi)$ from a $4$-dimensional space, does not
entail that they cannot be used to calculate the probability of a
transition from $\emptyset$ to $(\Sigma,\gamma,\phi)$.

This counter-argument is inconsistent with the principle that the
probability of a transition between two configurations is
determined by all the kinematically possible classical histories
that can interpolate between those configurations. Quantum
`tunnelling' occurs in non-relativistic quantum theory if there is
a transition which is not dynamically possible according to the
classical dynamical equations. However, quantum tunnelling in
non-relativistic quantum theory can only take place between two
configurations, $q_1$ and $q_2$, if there is a kinematically
possible classical history that interpolates between them. If
there is no such kinematically possible history, then even in
quantum theory, a transition between the two configurations is not
possible. For example, if $q_1$ and $q_2$ are points that belong
to disconnected regions of space, then a transition between $q_1$
and $q_2$ is impossible. Because no manifold $\Sigma$ can be
cobordant with the empty set, there are no kinematically possible
classical histories which interpolate between $\emptyset$ and
$(\Sigma,\gamma,\phi)$. Hence, there cannot be a quantum
transition between $\emptyset$ and $(\Sigma,\gamma,\phi)$. In
other words, quantum tunnelling between $\emptyset$ and
$(\Sigma,\gamma,\phi)$ is impossible.

\subsection{The Hartle-Hawking \textit{Ansatz}}

The most notorious application of Euclidean path-integral quantum
gravity to quantum cosmology is the paper of Hartle and Hawking
(1983). It is suggested here that the wave-function of the
universe $\Psi_0$ can be specified by Euclidean path-integration,
hence the Hartle-Hawking approach provides a meeting point between
the canonical approach and path-integral approach to quantum
gravity. The zero subscript indicates that Hartle-Hawking consider
this wave-function to be a type of `ground state', which normally
means a quantum state of minimum energy. Of all the possible
solutions to the Wheeler-DeWitt equation, it is suggested that the
`Euclidean' creation \textit{ex nihilo} path-integral generates
the correct solution. In elementary quantum theory, a
time-dependent quantum state-function, which satisfies the
Schrödinger equation, can be generated by a path-integral; here,
it is suggested that the time-independent state-function of
quantum gravity, which satisfies the Wheeler-DeWitt equation, can
also be generated by a path-integral. Hartle-Hawking include the
$3$-geometries $\gamma$ and the matter fields $\phi$ in the domain
of the wave-function, $\Psi_0(\gamma,\phi)$. One could also
include the $3$-topology $\Sigma$, although Hartle-Hawking
restrict their proposal to compact $3$-manifolds.

The Hartle-Hawking \textit{Ansatz} can be analysed into three
separate propositions.\footnote{It must be emphasised that Hartle
and Hawking made no such threefold distinction themselves.} The
first proposition is that the probability amplitudes
$K(\emptyset;\Sigma_f,\gamma_f,\phi_f)$ are the probability
amplitudes of creation \textit{ex nihilo}. The second proposition
is that these probability amplitudes provide the wave-function of
the universe $\Psi_0(\Sigma_f,\gamma_f,\phi_f)$. i.e.

$$
\Psi_0(\Sigma_f,\gamma_f,\phi_f) =
K(\emptyset;\Sigma_f,\gamma_f,\phi_f) \,.
$$

The third proposition is that
$K(\emptyset;\Sigma_f,\gamma_f,\phi_f)$ is specified by
path-integration over compact Riemannian $4$-geometries. Given the
intractability of the full path-integral, a weaker but more
plausible proposition can be substituted here:
$K(\emptyset;\Sigma_f,\gamma_f,\phi_f)$ is specified approximately
by a summation over select compact Riemannian $4$-geometries.
There are $2^3=8$ possible combinations for accepting or rejecting
these propositions. For example, one could agree that the
probability amplitudes $K(\emptyset;\Sigma_f,\gamma_f,\phi_f)$ are
equivalent with the wave-function of the universe, but one could
reject the proposal that these probability amplitudes are
generated by summation over compact Riemannian $4$-geometries. One
might attempt to use non-compact geometries and Lorentzian
geometries instead.

Conversely, one could agree that the probability amplitudes
$K(\emptyset;\Sigma_f,\gamma_f,\phi_f)$ are generated by summation
over compact Riemannian $4$-geometries, but one need not believe
that these amplitudes are equivalent with the wave-function of the
universe. Given that the wave-function of the universe is a
concept drawn from canonical quantum gravity, one could refuse to
grant that it has any connection with path-integral quantum
gravity.

Alternatively again, one could accept that the probability
amplitudes $K(\emptyset;\Sigma_f,\gamma_f,\phi_f)$ are equivalent
with the wave-function of the universe, and one could accept that
these amplitudes are determined by summation over compact
Riemannian $4$-geometries, but one could deny that these
probability amplitudes should be interpreted as creation
\textit{ex nihilo} probability amplitudes.

\hfill \break

There seems to be a degree of conceptual confusion in the original
expression of the Hartle-Hawking \textit{Ansatz}. For example,
consider the following passage: ``our proposal is that the sum
should be over compact geometries. This means that the Universe
does not have any boundaries in space or time (at least in the
Euclidean regime). There is thus no problem of boundary
conditions. One can interpret the functional integral over all
compact four-geometries bounded by a given three-geometry as
giving the amplitude for that three-geometry to arise from a zero
three-geometry, i.e. a single point. In other words, the ground
state is the amplitude for the Universe to appear from nothing,"
(Hartle and Hawking 1983, p2961).

This statement is open to a number of criticisms. Firstly, it is
entirely conventional in general relativistic cosmology to
represent the universe by a boundaryless differential manifold. It
is far from radical to suggest that the universe has no boundary
in space or time. Secondly, the concept of a compact $4$-manifold
is distinct from the concept of a boundaryless $4$-manifold. A
compact manifold may or may not possess a boundary. A boundaryless
manifold may be compact or non-compact. A manifold with boundary
may be compact or non-compact. By summing over compact
$4$-manifolds, one would exclude non-compact $4$-manifolds from
one's purview, but one would not exclude compact manifolds which
possess a boundary; the Hartle-Hawking proposal is to sum over
compact $4$-manifolds which possess a single $3$-dimensional
boundary component.

Thirdly, by moving from classical general relativity to
path-integral quantum gravity, one ceases to represent the
universe by a single space-time. It is, therefore, difficult to
understand in what sense it is `the Universe' which could be
bereft of boundary. In quantum cosmology, the universe is
represented by a wave-function, not a manifold.

Fourthly, the so-called `Euclidean regime' is a distinct concept
from summation over compact $4$-geometries. One could propose
summation over compact $4$-geometries without proposing that the
$4$-geometries must be Riemannian (`Euclidean').

There appear to be two types of boundary conditions at work in the
Hartle-Hawking \textit{Ansatz}. There are boundary conditions on
the hypothetical wave-function of the universe, and there are
boundary conditions on the individual $4$-geometries in the
summation. The claim in the above excerpt that there is no problem
with boundary conditions, implies that the boundary conditions
referred to at this juncture are boundary conditions on the
4-geometries in the summation, not boundary conditions on the
wave-function. It is only for compact $4$-geometries that the
action is guaranteed to be finite. If one were to permit
non-compact $4$-geometries, one would have to impose boundary
conditions to ensure that the action integral of such
$4$-geometries did not diverge. Hartle-Hawking propose that the
wave-function be obtained by summation over compact
$4$-geometries, which need no spatial boundary conditions. This is
the proposed boundary condition on the wave-function.

The confusion created by the Hartle-Hawking \textit{Ansatz}, and
by the decision to name it the `no-boundary' boundary condition,
is typified by the account given by Kolb and Turner: ``because a
compact manifold has no boundaries, this proposal is referred to
as the `no-boundary' boundary condition," (Kolb and Turner 1990,
p462). To reiterate, a compact manifold can have a boundary, and a
non-compact manifold need not have a boundary.

The ambiguity of the phrase `boundary conditions', is used by
Hawking in his well-known dictum that ``the boundary conditions of
the Universe are that it has no boundary." Hawking has stated that
``if spacetime is indeed finite but without boundary or edge...it
would mean that we could describe the Universe by a mathematical
model which was determined completely by the laws of science
alone; they would not have to be supplemented by boundary
conditions," (Hawking 1989, p69). A statement like this suppresses
the fact that in path-integral quantum gravity, one no longer
represents the universe by an individual space-time; one deals
with summation over multiple space-times.

The assertion that the laws of science would ``not have to be
supplemented by boundary conditions" is even more unfathomable
because Hawking freely admits that the Euclidean no-boundary
proposal ``is simply a proposal for the boundary conditions of the
Universe," (Hawking 1989, p68). Hawking must know that the
Wheeler-DeWitt equation is a proposed `law of science' which has
many possible solutions, and to select a particular solution, one
needs to specify boundary conditions. Hawking's misleading claims
for the `no-boundary' proposal have been widely disseminated.
Barrow, for example, claims that the Hartle-Hawking
\textit{Ansatz} ``removes the conventional dualism between laws
and initial conditions," (Barrow 1991, p67).

Returning to the excerpt from the 1983 paper, Hartle-Hawking
interpret their ground state wave-function as giving the amplitude
for any $3$-geometry ``to arise from a zero three-geometry, i.e. a
single point. In other words, the ground state is the amplitude
for the Universe to appear from nothing." Hartle-Hawking introduce
three distinct concepts here, and treat them as if they are
equivalent. First of all they refer to a ``zero three-geometry",
then they refer to a ``single point", then they refer to
``nothing". The number zero is a \textit{bona fide} element of the
set of real numbers, and is quite distinct from nothing, the empty
set $\emptyset$. Moreover, it is not clear what Hartle-Hawking
mean by a ``zero three-geometry". A single point is sometimes
considered by mathematicians to be a zero-dimensional manifold,
but such an object cannot have any geometry, never mind a ``zero
three-geometry". Furthermore, single points never appear in the
summations under consideration. The proposed summations are over
manifolds which have no initial boundary, so one is dealing with
the empty set $(\Sigma_i,\gamma_i,\phi_i) = \emptyset$, not a
single point, and not some mythical ``zero three-geometry".

\subsection{The WKB and steepest-descent approximations}

In papers such as Halliwell (1991), Halliwell and Hartle (1990),
Gibbons and Hartle (1990), the `Euclidean' creation \textit{ex
nihilo} proposal, (the Hartle-Hawking \emph{Ansatz}), developed
into the following `sum-over-histories':\footnote{We shall use
square brackets hereafter to enclose arguments which are functions
or fields on manifolds.}

$$
\Psi_0[\Sigma,\gamma,\phi] = \sum_\mathcal{M} \nu(\mathcal{M})
\int e^{-A_E[\mathcal{M},g,\Phi]/\hbar} \, d\mu[g,\Phi] \, .
$$

The sum here is understood to be over 4-manifolds $\mathcal{M}$
which are bounded by $\Sigma$. $\nu(\mathcal{M})$ is a weight that
one assigns to each such 4-manifold. For each 4-manifold
$\mathcal{M}$ bounded by $\Sigma$, the integral is over compact
Riemannian 4-geometries and matter field histories $(g,\Phi)$ on
$\mathcal{M}$, which induce $(\gamma,\phi)$ on $\Sigma$. By
summing over 4-manifolds, it is tacitly assumed that there are
only a countable number of 4-manifolds bounded by $\Sigma$. As
usual, $\Sigma$ is only considered to be a compact 3-manifold.

The approach taken in papers such as those listed above is to
reject the assumption that the `sum-over-histories' should be
taken over all the histories $(\mathcal{M},g,\Phi)$ which are
bounded by $(\Sigma,\gamma,\phi)$. Instead, the manifolds that one
should sum over are to be a matter of debate; the weight of each
such 4-manifold is to be determined; one restricts the domain of
integration to a `contour' of integration, and the particular
contour chosen is to be a matter of debate. The restriction of the
domain of integration is intended to find a way of making the
path-integral convergent, and the measure upon the domain of
integration is also considered to be a matter of debate. Only some
combinations of these choices, it is argued, will lead to a
convergent `sum-over-histories'. Moreover, different combinations
of these choices will lead to different wave-functions.

In practice, the `Euclidean' creation \textit{ex nihilo} proposal
for the wave-function, has only been applied to mini-superspace
models, and even then, the `Euclidean' path-integrals have not
been calculated. Instead, the so-called `steepest-descent'
approximation to the path-integral has been used to obtain a WKB
approximation to the wave-function, or a sum of such WKB
wave-functions.

The phase of a WKB wave-function approximately satisfies the
classical Hamilton-Jacobi equation, hence the WKB approximation is
often referred to as the `semi-classical' approximation. In
quantum cosmology, however, this can confuse matters because, as
we shall see, there is another, more specific sense in which the
term `semi-classical' is used.

The difference between `oscillatory' and `exponential' WKB
wave-functions has interpretational significance in quantum
cosmology, hence a digression to explain the difference is
worthwhile. If we abstract momentarily from the context of quantum
cosmology, the WKB approximation is used to obtain an approximate
wave-function $\Psi(x)$ under the conditions where the de Broglie
wavelength function $\lambda(x)$ does not change significantly
over the distance of one wave-length. For a system of energy $E$,
with a potential $V(x)$,

$$
\lambda(x) = \frac{2\pi\hbar}{k(x)} = \frac{2\pi\hbar}{\sqrt{2m[E
- V(x)]}} \, .
$$ Given that $\lambda(x)$ only changes by virtue of a change in the
potential $V(x)$, the WKB approximation is valid wherever the
potential does not change significantly over the distance of one
wavelength.

The domain of a WKB wave-function can be divided into a
classically-permitted region, where $E > V(x)$, and a
classically-forbidden region, where $E < V(x)$. For the first
order WKB approximation, in the classically permitted region the
wave-function will be `oscillatory':

$$
\Psi(x) = \frac{a_1}{\sqrt{k(x)}}\exp \left[(i/\hbar) \int_{x_0}^x
k(x') dx' \right ] + \frac{a_2}{\sqrt{k(x)}}\exp \left [(-i/\hbar)
\int_{x_0}^x k(x') dx' \right] \, .
$$

In the classically forbidden region the wave-function will be
exponential:

$$
\Psi(x) = \frac{a_1}{\sqrt{|k(x)|}}\exp \left[(1/\hbar)
\int_{x_0}^x |k(x')| dx'\right] + \frac{a_2}{\sqrt{|k(x)|}}\exp
\left [(-1/\hbar) \int_{x_0}^x |k(x')| dx'\right] \, .
$$

In the classically forbidden region, $k(x)=\sqrt{2m[E - V(x)]}$ is
an imaginary number, hence the difference in the expressions.
Given that $\lambda(x) = (2\pi \hbar/k(x))$, the de Broglie
wavelength will also be imaginary in the forbidden region.

The classical turning points are the regions in which $|E - V(x)|$
is small. In these regions, $k(x)$ becomes very small and
$\lambda(x)$ becomes very large. Hence, in the regions near the
classical turning points, the change in $\lambda(x)$ can be
significant over the distance of a wavelength. i.e. $V(x)$ can
vary significantly over the distance of a wavelength. Hence, the
WKB approximation is not valid in the regions near the classical
turning points.

The WKB approximation is often defined to be valid for a
wave-function $\Psi(x) = C(x)e^{iS(x)}$ wherever the phase $S(x)$
is rapidly varying relative to the modulus $C(x)$. At first sight,
this might seem to suggest that the WKB approximation is invalid
in the classically forbidden, exponential regions, where the
wave-function is real-valued, and therefore of constant (zero)
phase but varying modulus. However, even though $\Psi(x)$ is
real-valued in the forbidden region, one can express the
real-valued exponential $\exp [(\pm 1/\hbar)\int_{x_0}^x |k(x')|
dx']$ in the form

$$
\exp [\pm i S(x)] = \exp \left [(\pm i/\hbar)\int_{x_0}^x k(x')
dx' \right ]\, ,$$ and it is this imaginary-valued phase $S(x) =
(\pm 1/\hbar)\int_{x_0}^x k(x') dx'$ which is rapidly-varying. The
phase-change per unit length in the oscillatory and exponential
regions is simply $k(x)/\hbar$, hence small values for
$\lambda(x)$ mean large values for $k(x)$, and therefore a
rapidly-varying phase. The only difference in the forbidden region
is that $k(x)$ is imaginary.

A WKB wave-function is defined analogously in quantum cosmology.
The regions of the configuration space in which the wave-function
can be given a WKB approximation are those regions in which the
phase is rapidly varying with respect to the modulus, entailing
that the phase \emph{approximately} satisfies the classical
time-independent Hamilton-Jacobi equation of canonical general
relativity, (Isham 1992b, p79).

Now, the steepest-descent approximation to a path-integral in
quantum theory is the proposition that in some regions of
configuration space there is no need to calculate the entire
path-integral. Instead, one need only consider the actions of
paths which are classical solutions. In terms of quantum
cosmology, if a spatial configuration $(\Sigma,\gamma,\phi)$ lies
in a region where the steepest-descent approximation to the
path-integral is valid, then one need not consider all
4-dimensional Riemannian histories bounded by
$(\Sigma,\gamma,\phi)$. Instead, one need only consider those
Riemannian 4-geometries which are saddle points of the action.

A solution of the classical Einstein field equations is a
stationary point of the action, and a saddle point is a special
kind of stationary point,\footnote{A saddle point is a stationary
point which is not an extremum.} hence a saddle point is a special
type of classical solution. Of all the Riemannian 4-geometries
bounded by $(\Sigma,\gamma,\phi)$, the claim is that one need only
consider those which are saddle point solutions of the classical
Einstein field equations. It is assumed, or reasoned, that the
contributions from other Riemannian 4-geometries are either
negligible, or cancel out.

The simplest version of the steepest-descent approximation
proposes that to each $(\Sigma,\gamma,\phi)$, there is some
Riemannian 4-geometry $(\mathcal{M},g,\Phi)$, with boundary
$(\Sigma,\gamma,\phi)$, which is a \emph{dominant} saddle point of
the action functional $A_E$. Where this approximation is valid,
the wave-function then has the form\footnote{$\hbar$ is omitted
hereafter to avoid unnecessary clutter.}

$$
\Psi_0 \sim \nu(\mathcal{M})
\Delta_{WKB}[\Sigma,\gamma,\phi;\mathcal{M},g,\Phi]
e^{-A_E[\Sigma,\gamma,\phi;\mathcal{M},g,\Phi]}.
$$ Notice the presence of the so-called WKB pre-factor
$\Delta_{WKB}$.

A less simple version of the approximation proposes that amongst
the 4-geometries with boundary $(\Sigma,\gamma,\phi)$, there can
be multiple saddle points of the action,
$\{(\mathcal{M}_i,g_i,\Phi_i): i = 1,2,...\}$, with no single
dominant contribution. The wave-function then has the form

$$
\Psi_0 \sim \sum_i \nu(\mathcal{M}_i)
\Delta_{WKB}[\Sigma,\gamma,\phi;\mathcal{M}_i,g_i,\Phi_i]
e^{-A_E[\Sigma,\gamma,\phi;\mathcal{M}_i,g_i,\Phi_i]}.
$$

The proponents of the steepest-descent approximation in quantum
cosmology claim that dominant contributions to the path-integral
come from \emph{complex} 4-geometries which are saddle points of
the action. Halliwell states that ``one generally finds that the
dominating saddle-points are four-metrics that are not real
Euclidean, or real Lorentzian, but complex, with complex action,"
(Halliwell 1991, p185). He also asserts that ``it appears to be
most commonly the case for generic boundary data that no real
Euclidean solution exists, and the only solutions are complex,"
(Halliwell 1991, p184).

Halliwell and Hartle state that ``a semi-classical approximation
to $\Psi_0$$\ldots$ arises when, in the steepest descent
approximation to the functional integral, the dominating saddle
points are complex," (1990, p1817). This is the second sense in
which the wave-function can be `semi-classical'. Used in this
context, the term is not meant to be synonymous with the WKB
approximation, but to indicate that there is a sense in which
classical space-times can be recovered from the wave-function, as
we shall see below.

Halliwell points out that the `Euclidean' gravitational action
$A_E$ of real Riemannian (`Euclidean') 4-geometries is unbounded
from below. ``This means that the path-integral will not converge
if one integrates over real Euclidean metrics," he asserts.
``Convergence is achieved only by integrating along a complex
contour in the space of complex four-metrics," (Halliwell 1991,
p172). For a steepest-descent approximation to a path-integral to
be valid, there must be a contour of integration which passes
through the relevant stationary points, and whose contributions
decline rapidly away from those stationary points. i.e. there must
be a steepest-descent contour, (Butterfield and Isham 1999, p55).

The practitioners of quantum cosmology claim that complex
geometries are vital to recovering the notion of classical
Lorentzian space-time. To reiterate, in those regions of
configuration space where the wave-function can be given a WKB
approximation, the wave-function can be either exponential
$C\,exp(-S)$, or oscillatory $C\,exp(iS)$, (Halliwell 1991, p181).
It is in the regions where the wave-function is oscillatory that
the notion of a classical Lorentzian space-time can be recovered.
Consider again the steepest-descent approximation for a single
dominant saddle point:

$$
\Psi_0 \sim \nu(\mathcal{M})
\Delta_{WKB}[\Sigma,\gamma,\phi;\mathcal{M},g,\Phi]
e^{-A_E[\Sigma,\gamma,\phi;\mathcal{M},g,\Phi]} \, .
$$

If $A_E$ is real, then the wave-function is clearly exponential
$C\,exp(-S)$. If, however, $A_E$ is complex, then the
wave-function will be oscillatory. If $A_E$ is a complex number,
as it is for a complex saddle point, then $A_E = Re(A_E) + i
Im(A_E)$, and one can factorize $exp(-A_E)$ as follows:

$$\eqalign{
e^{-A_E} &= e^{-Re(A_E)-i Im(A_E)} \cr &= e^{-Re(A_E)} e^{-i
Im(A_E)} \, .}$$ In this event the wave-function
$\Psi_0[\Sigma,\gamma,\phi]$ can be written as

$$
\Psi_0 \sim \nu(\mathcal{M})
\Delta_{WKB}[\Sigma,\gamma,\phi;\mathcal{M},g,\Phi]
e^{-Re(A_E[\Sigma,\gamma,\phi;\mathcal{M},g,\Phi])} e^{-i Im(
A_E[\Sigma,\gamma,\phi;\mathcal{M},g,\Phi])} \, .
$$

This wave-function clearly has the oscillatory form
$\Psi_0[\Sigma,\gamma,\phi] =
C[\Sigma,\gamma,\phi]e^{iS[\Sigma,\gamma,\phi]}$. The modulus
$C[\Sigma,\gamma,\phi]$ is given by

$$
C[\Sigma,\gamma,\phi] = \nu(\mathcal{M})
\Delta_{WKB}[\Sigma,\gamma,\phi;\mathcal{M},g,\Phi]
e^{-Re(A_E[\Sigma,\gamma,\phi;\mathcal{M},g,\Phi])}
$$ In this case, the phase of the wave-function is determined by the
imaginary part $-Im(A_E[\Sigma,\gamma,\phi;\mathcal{M},g,\Phi])$
of the action, and the real part of the action contributes the
factor

$$
e^{-Re(A_E[\Sigma,\gamma,\phi;\mathcal{M},g,\Phi])}
$$ to the modulus. Being exponential, this factor can dominate the
modulus, hence the square modulus $exp(-2Re(A_E))$ is often taken
to provide a probability distribution. The smaller the real part
of the action, the greater the contribution.

According to Halliwell, an oscillatory WKB wave-function is peaked
about a set of Lorentzian solutions of the classical equations.
The `classical trajectories' are defined to be the integral curves
of the vector field $\nabla S$, the gradient of the phase of the
wave-function. The wave-function is claimed to be peaked, not
about a single classical solution, but about a set of classical
solutions. The integral curves of $\nabla S$ constitute a
congruence of the subset of the configuration space in which the
wave-function can be approximated by an oscillatory WKB
wave-function.

Halliwell claims that the squared-modulus $| C | ^2$ is constant
along each classical trajectory, and therefore provides a
probability measure on the classical trajectories.

\subsection{Signature change space-times and intrinsic time}

To illustrate the emergence of classical Lorentzian paths in those
regions of the configuration space in which an oscillatory WKB
wave-function is valid, let us consider the case of signature
change space-times. Those signature-change space-times
$\mathcal{M}$ relevant to the Hartle-Hawking \emph{Ansatz} consist
of a 4-dimensional region of compact Riemannian geometry
$\mathcal{M}_R$, in which there is no time, an adjoining
4-dimensional region of Lorentzian space-time $\mathcal{M}_L$, and
a compact 3-dimensional signature-changing hypersurface $\Sigma$,
which separates the two regions, so that (Gibbons and Hartle 1990,
p2460)

$$
\mathcal{M} = \mathcal{M}_L \cup \mathcal{M}_R
$$
$$
\partial \mathcal{M}_L = \Sigma = \partial \mathcal{M}_R \, .
$$

One considers a Riemannian metric on $\mathcal{M}_R$ and a
Lorentzian metric on $\mathcal{M}_L$, which are such that they
induce the same spacelike geometry on the 3-manifold $\Sigma$.

Although such a signature-change space-time is not a complex
4-geometry, it does define a complex action, with the action of
the Riemannian region providing the real component, and the action
of the Lorentzian region providing the imaginary component,
(Gibbons and Hartle 1990, p2460).

Suppose that one has a signature-change space-time which is a
saddle point of the complex action, and suppose that the
Lorentzian region can be foliated by a one-parameter family of
spacelike hypersurfaces. Each space-like slice
$(\Sigma,\gamma,\phi)$ can be treated as the boundary of a
signature-change space-time if one removes the Lorentzian region
to the future of that slice. The signature-change space-time
consists of the prior region of Lorentzian space-time, and the
entire region of Riemannian space. Assuming that the complex
action $A_E$ of this truncated signature-change space-time
provides the dominant saddle-point contribution to the
wave-function value $\Psi_0[\Sigma,\gamma,\phi]$ in the
steepest-descent approximation to the Hartle-Hawking
path-integral, one can set $\Psi_0 \sim e^{-A_E}$. The actions of
the 4-geometries bounded by the slices $(\Sigma,\gamma,\phi)$ in
the Lorentzian region differ only by the value of their imaginary
component. The real component is determined by the action of the
Riemannian region, and this is common to all the slices in the
Lorentzian region, hence the real component does not vary. The
real component determines the modulus $C[\Sigma,\gamma,\phi]$ of
the wave-function, hence the modulus of the wave-function does not
vary for the slices in the Lorentzian region. The imaginary
component determines the phase of the wave-function, hence the
phase of the wave-function varies amongst the slices in the
Lorentzian region. The gradient of the phase, $\nabla S$,
therefore gives the classical Lorentzian paths in that part of the
configuration space in which the wave-function is determined by
dominant signature-change saddle points, part of the region in
which the wave-function is said to be oscillatory.

In those regions of the configuration space where the
wave-function is exponential, one does \emph{not} have a
congruence of classical `Euclidean' paths, (Halliwell 1991, p182).
Suppose that one has a real 4-dimensional compact Riemannian
geometry, which is a saddle-point solution of the classical field
equations. Suppose that this geometry can be foliated by a
one-parameter family of hypersurfaces which are, of necessity,
spacelike themselves, and suppose that one chooses an ordering for
these slices. Each slice $(\Sigma,\gamma,\phi)$ can be treated as
the boundary of a compact Riemannian 4-geometry if one removes the
region which is ordered to be `greater than'
$(\Sigma,\gamma,\phi)$. Moreover, such an operation renders
$(\Sigma,\gamma,\phi)$ as the boundary of a saddle-point solution
of the classical equations. Assuming the steepest-descent
approximation to the Hartle-Hawking path-integral gives the
wave-function value $\Psi_0[\Sigma,\gamma,\phi]$, and assuming
that the action $A_E$ of this truncated Riemannian region provides
the dominant saddle-point contribution, one can set $\Psi_0 \sim
e^{-A_E}$. For successive slices $(\Sigma,\gamma,\phi)$ in the
Riemannian 4-geometry, the size of the truncated region grows, and
the real-valued action and wave-function vary accordingly. Hence,
the path in configuration space consisting of successive slices of
the Riemannian 4-geometry is assigned complex numbers of constant
(zero) phase and varying modulus. The integral curves of $\nabla
S$ do not correspond to these paths, and the wave-function is not
peaked over such paths in configuration space.

This nice clean distinction between regions of the configuration
space breaks down if one considers configurations which can be
embedded in both a foliation of the Lorentzian region in a
signature-change saddle-point space-time, and in a foliation of a
saddle-point Riemannian 4-geometry.

\hfill \break

It is often said that classical space-time must be a prediction of
quantum cosmology in the `late' universe. This means that the
oscillatory WKB approximation must be valid in the part of the
configuration space which contains the large 3-geometries (i.e.
large scale-factors), because large geometries correspond to the
universe that we presently inhabit. Not only is the concept of
Lorentzian space-time lost in those regions of the configuration
space in which the oscillatory WKB approximation is invalid, but
the notion of time itself seems to be lost in those regions.

In quantum mechanics in general, wherever the oscillatory WKB
approximation is valid, one can interpret the wave-function to
describe a classical statistical ensemble. In this sense, wherever
the wave-function of quantum gravity is given by an oscillatory
WKB approximation, one does indeed recover a family of Lorentzian
space-times. However, in the particular case of quantum cosmology,
this poses an interpretational problem because it implies a
statistical ensemble of universes. It appears that even where the
wave-function is `semi-classical', its interpretation requires one
to accept that there are many universes, each of which follows one
of the classical trajectories.

The recovery of classical space-times faces a problem when there
is more than one stationary point in the steepest-descent
approximation to the wave-function. The wave-function then has the
form

$$
\Psi_0[\Sigma,\gamma,\phi] \sim \sum_i C_i[\Sigma,\gamma,\phi]
e^{i S_i[\Sigma,\gamma,\phi]}
$$

As a consequence, there is an entire family of different
congruences, each determined by $\nabla S_i$. One no longer has a
unique family of classical trajectories. The interpretational
difficulties are therefore magnified. Even if one postulates the
existence of many universe, the wave-function in this case would
seem to describe those universes to be in a state of quantum
superposition.\footnote{The reader should be aware that physicists
are fond of something called the decoherent histories
interpretation, which purportedly explains how this is consistent
with our observation of an individual classical space-time. See
Halliwell (1989).}

\hfill \break

Kossowski and Kriele (1994a, p115 and 1994b, p297) suggest that
the `Euclidean' creation \textit{ex nihilo} proposal reduces in
the classical limit to a signature-changing space-time. They state
that ``our point of view is that the path-integral argument of
Hartle and Hawking gives initial conditions for the classical
Einstein equation at the [signature change hypersurface],"
(Kossowski and Kriele, 1994a, p116). On the previous page, (p115),
they assert that Hartle and Hawking are able to ``calculate rather
than assume the initial state for the Lorentzian part of the
universe $\ldots$ they obtain an initial state at [the signature
change hypersurface] by path integration over all Riemannian
metrics (defined on the Riemannian region)." If the oscillatory
region of the configuration space can be foliated by the
congruence $\nabla S$ of classical Lorentzian space-times, and the
modulus is constant along each such path, then the wave-function
on the boundary between the oscillatory region and the exponential
region would effectively determine a probability distribution
across a family of initial configurations for classical Lorentzian
space-times.

Let us suppose that we have fixed a 4-dimensional manifold
$\mathcal{M} = \mathcal{M}_L \cup \mathcal{M}_R$ as above.
Presumably, Kossowski and Kriele would wish to consider, for each
possible pair $(\gamma,\phi)$ on $\Sigma$, a path-integral over
all possible Riemannian 4-geometries and matter field histories
$(g,\Phi)$ on $\mathcal{M}_R$ which induce $(\gamma,\phi)$ on
$\Sigma$. According to the Hartle-Hawking proposal, this would
yield a quantum state-function $\Psi_0[\gamma,\phi]$, whose domain
is the set of all possible 3-metrics and matter fields on the
hypersurface $\Sigma$. Kossowski and Kriele propose that this
state function should be considered as ``the initial state for the
Lorentzian part of the universe." Unfortunately, the quantum
state-function $\Psi_0[\gamma,\phi]$ does not constitute initial
conditions for the classical Einstein equation, and
$|\Psi_0[\gamma,\phi]|^2$ does not even constitute a probability
distribution over the initial conditions for the Einstein
equation. Recall that initial conditions in a classical theory do
not merely consist of a configuration, but also a rate of change
of configuration. Initial conditions for the Einstein equation on
a hypersurface $\Sigma$ consist of a 3-metric, the extrinsic
curvature tensor or conjugate momentum tensor field, the matter
fields, and the first order matter field time derivatives. At
best, the Hartle-Hawking wave-function could provide quantum
initial conditions, rather than the classical initial conditions
suggested by Kossowski and Kriele. To reconcile this with the
apparent time-independence of the wave-function, it is necessary
to invoke the notion of `intrinsic' time.

The idea here is that time can be found in the domain of the
wave-function. A genuine configuration space in canonical quantum
gravity is infinite-dimensional; there will be an infinite number
of degrees of freedom. Intrinsic time advocates suggest that one
can split the degrees of freedom into those which are `physical',
and those which are `non-physical'. The physical degrees of
freedom are sufficient to pin down the configuration, whilst the
non-physical are redundant degrees of freedom, which purportedly
contain information about intrinsic time.

Isham (1988, p396) argues that since the degrees of freedom
include an internal definition of time, it would be incorrect to
add an external time label to the state function $\Psi$. Instead,
the internal time is treated as a function $T[\Sigma,\gamma,\phi]$
of the configuration, and $\Psi[\Sigma,\gamma,\phi]$ gives the
probability amplitude of the physical configuration
$(\Sigma,\gamma,\phi)_{phys}$ at the internal time
$T[\Sigma,\gamma,\phi]$.

One could presumably fix the physical degrees of freedom, but
allow the internal time to vary; the probability amplitude of a
physical configuration would vary with internal time. One could
also, presumably, fix the value of the internal time, and consider
all the possible physical configurations at that value of the
internal time. The square-modulus of the wave-function would then
provide a probability distribution over all the possible physical
configurations at that internal time. By allowing the internal
time to vary, one would have a varying probability distribution
over the possible physical configurations. One could write the
wave-function as

$$
\Psi[\Sigma,\gamma,\phi]= \Psi[(\Sigma,\gamma,\phi)_{phys},T] =
\Psi_T[\Sigma,\gamma,\phi]_{phys} \, .
$$

There are supposedly many different choices of internal time. The
Wheeler-DeWitt equation purportedly governs the time-dependence of
the wave-function for any choice of internal time. If one were to
specify $\Psi_0[\Sigma,\gamma,\phi]_{phys}$ at some internal time
$T = 0$, then the Wheeler-DeWitt equation would purportedly
determine $\Psi_T[\Sigma,\gamma,\phi]_{phys}$ at any other value
$T$ of internal time.

The notion of intrinsic time can also be applied to path-integral
quantum gravity. One interprets the transition amplitude

$$
K(\Sigma_i,\gamma_i,\phi_i;\Sigma_f,\gamma_f,\phi_f) \, ,
$$ as the amplitude of a transition from the physical configuration
$(\Sigma_i,\gamma_i,\phi_i)_{phys}$ at the internal time
$T[\Sigma_i,\gamma_i,\phi_i]$, to the physical configuration
$(\Sigma_f,\gamma_f,\phi_f)_{phys}$ at the internal time
$T[\Sigma_f,\gamma_f,\phi_f]$.

The idea of intrinsic time is, however, difficult to implement in
practice, and existing attempts use mini-superspace models. The
concept of (internal) time may have a limited domain of validity.

G.F.R. Ellis asserts that the Hartle-Hawking \textit{Ansatz} is
``a scheme whereby the origin of the Universe is separated from
the issue of the origin of time," (Ellis 1995, p326). This is a
dubious interpretation. Recall that part two of the Hartle-Hawking
\emph{Ansatz} is that the probability amplitude of
$(\Sigma,\gamma,\phi)$ being created from nothing, is given by a
path-integral over compact Riemannian 4-geometries which are
bounded by $(\Sigma,\gamma,\phi)$. It could then be suggested that
once a universe has been created \textit{ex nihilo}, it evolves as
a Lorentzian space-time thereafter. This is a distinct proposal,
and not one made by Hartle and Hawking, but let us consider it for
the sake of argument. Even then, one need not accept Ellis'
interpretation that the origin of time is separate from the origin
of the universe. One could suggest that the Riemannian 4-manifolds
only have a part to play in the creation \textit{ex nihilo}
calculations, not in any actual processes, hence there would be no
actual signature change process. One would merely integrate over
Riemannian 4-manifolds to find the creation \textit{ex nihilo}
probabilities. There would be creation from nothing, and
Lorentzian space-time thereafter, with no intermediate Riemannian
geometry. In this case, the creation of a universe would coincide
with the creation of time, contrary to Ellis' suggestion.

Isham recognizes that one need not ascribe physical status to the
Riemannian geometries used in the Hartle-Hawking definition of the
wave-function $\Psi_0$. He recognizes that a `` `phenomenological'
four-dimensional (Lorentzian) space-time that is reconstructed
from the canonical state $\Psi_0$ is not necessarily related,
either metrically or topologically, to any four-dimensional
manifold that happens to be used in the construction of the state.
Indeed, if the concept of `time' is only semi-classical, it is
incorrect to talk at all about a four-dimensional manifold at the
quantum level," (Isham 1991b, p356).

It is the use of signature-change saddle points in the
steepest-descent approximation to the wave-function
$\Psi_0[\Sigma,\gamma,\phi]$, rather than the use of the genuine
`Euclidean' path-integral, which has inspired some authors to draw
a line between the creation of a universe, and the origin of time.
Gibbons and Hartle (1990, p2459), for example, consider a
signature change solution of the classical Einstein equation, to
be a ``tunneling solution," . They state that such ``tunneling
solutions describe the universe `tunneling from nothing', and are
the dominant contributors to the semiclassical approximations to
the `no-boundary' proposal," (1990, p2460).

To describe a signature change solution of the classical Einstein
equation as a tunnelling solution, brings tunnelling down to the
level of classical theory, when it should be exclusively a quantum
phenomenon. Signature change should not, in itself, be considered
as an occurrence of tunnelling.

Another difficulty with the introduction of signature change
space-times, is that the interpretation of the probability
amplitude $\Psi_0[\Sigma,\gamma,\phi]$ assigned to a triple
$(\Sigma,\gamma,\phi)$ becomes ambiguous. Is
$\Psi_0[\Sigma,\gamma,\phi]$ the probability amplitude that
$(\Sigma,\gamma,\phi)$ be created from nothing, or is it the
probability amplitude of $(\Sigma,\gamma,\phi)$ being the initial
configuration of the Lorentzian region of the universe? Could it
even be both? Prima facie, the creation of $(\Sigma,\gamma,\phi)$
from nothing would seem to require a direct transition from
$\emptyset$ to $(\Sigma,\gamma,\phi)$. When $(\Sigma,\gamma,\phi)$
is the boundary of a 4-dimensional Riemannian region, it is
difficult to interpret it to have been created from nothing. Only
if one interprets the Riemannian 4-geometries bounded by
$(\Sigma,\gamma,\phi)$ as calculational fictions, could one
maintain the creation \textit{ex nihilo} interpretation.

One could argue that an individual signature-changing space-time
should be construed merely as a classical version of the quantum
tunnelling of quantum cosmology. But this then contradicts the
idea that the Riemannian region is classically forbidden, as seen
in the Hartle-Hawking de Sitter mini-superspace model that we will
encounter in Section 4. Is signature change part of quantization,
or is it a preparation for quantization?

If the wave-function of the universe is interpreted
epistemologically, so that it is thought to provide merely an
incomplete description, then one can interpret the wave-function
to provide a statistical description of an ensemble of universes.
If one interprets the probabilities of the wave-function $\Psi_0$
epistemologically, then one could conceivably assert that
individual signature change space-times exist in the statistical
ensemble. If, however, one interprets the wave-function and its
probabilities ontologically, then what actually exists would be
the wave-function $\Psi_0$. Individual signature-change
space-times would not exist.

\section{Mini-superspace quantum cosmology}

In an effort to make the process of finding solutions to the
Wheeler-DeWitt more tractable, `mini-superspace' models were
employed in quantum cosmology. Such models fix all but a finite
number of degrees of freedom before quantization. The intention in
this section is to review, clarify, and critically analyse
mini-superspace quantum cosmology. In particular, the claim that
Vilenkin's `tunnelling boundary condition' provides the
probability of creating a universe from nothing, will be subjected
to critical scrutiny.

To reiterate, in canonical general relativity, expressed in terms
of the `traditional' variables, the set of all possible
geometrical configurations of the spatial universe corresponds to
the set of all 3-dimensional Riemannian manifolds
$(\Sigma,\gamma)$. This set of all 3-dimensional Riemannian
geometries is a disconnected topological space. Each component of
the disconnected space corresponds to the set
$\mathscr{C}(\Sigma)$ of all Riemannian metric tensors upon a
fixed 3-manifold $\Sigma$.

Although $\mathscr{C}(\Sigma)$ is referred to as a configuration
space, there exist distinct elements of $\mathscr{C}(\Sigma)$
which are isometric. This can be understood by the action of
$Diff(\Sigma)$, the diffeomorphism group of $\Sigma$, upon the
space of metrics $\mathscr{C}(\Sigma)$. A diffeomorphism
$\phi:\Sigma \rightarrow \Sigma$ maps a metric $h \in
\mathscr{C}(\Sigma)$ to another metric $h'$ by pullback, $h' =
\phi^*h$. That is, $h'_p(v,w) = h_{\phi(p)}(\phi_*(v),\phi_*(w))$
at each point $p \in \Sigma$, and for each pair of vectors $v,w
\in T_p\Sigma$.

The orbits of the action of $Diff(\Sigma)$ are the isometry
equivalence classes of Riemannian metric tensor fields on
$\Sigma$. Hence, one considers the quotient $\mathcal{S}(\Sigma) =
\mathscr{C}(\Sigma)/Diff(\Sigma)$ to be the set of all possible
intrinsic Riemannian geometries of $\Sigma$. $\mathcal{S}(\Sigma)$
is known as the superspace of the 3-manifold $\Sigma$.

Whilst a wave-function $\Psi$ on the configuration space
$\mathscr{C}(\Sigma)$ must satisfy both the Wheeler-DeWitt
equation, and then additional constraint equations to ensure it is
invariant under diffeomorphisms of $\Sigma$, a wave-function on
superspace $\mathcal{S}(\Sigma)$ need merely satisfy the
Wheeler-DeWitt equation.

In the absence of matter, a true wave-function $\Psi[\gamma]$ of
the universe, in the traditional variables configuration
representation, would be a complex-valued functional upon the
entire disconnected space of possible 3-geometries. However, in
most of the existing literature on quantum cosmology, it is
conventional to tacitly restrict the topological degrees of
freedom; in particular, a compact and orientable 3-manifold
$\Sigma$ is fixed from the outset. Typically, the three-sphere
$S^3$ is chosen.

Even by fixing the topological degrees of freedom, however, the
set of all Riemannian 3-geometries and matter field configurations
on $\Sigma$ is still infinite-dimensional. Thus, even by omitting
the 3-topology as an argument of the wave-function, the latter
will still be a function $\Psi[\gamma,\phi]$ on an
infinite-dimensional manifold. In general, it is difficult to
solve differential equations on an infinite-dimensional manifold,
hence it is very difficult to find any solutions of the
Wheeler-DeWitt equation, and even more difficult to select a
unique solution which satisfies some `boundary' conditions.

Thus, in an effort to make things more tractable, mini-superspace
models were employed in quantum cosmology. In these models,
symmetries such as homogeneity and isotropy were imposed, and all
but a finite number of degrees of freedom were frozen
\emph{before} quantization. Hence, the superspace of such models
is a finite-dimensional submanifold of the full superspace, and
efforts can be made to find solutions of the Wheeler-DeWitt
equation restricted to such finite-dimensional
domains.\footnote{In Bojowald's recent application of loop quantum
gravity to quantum cosmology, he quantizes the kinematics of the
full theory, and then seeks quantum states which correspond, in
some sense, to homogeneous and isotropic space. See Ashtekar
(2002).}

To demonstrate the mini-superspace technique in the traditional
variables, we shall begin by considering a well-known model with
only one degree of freedom, (Kolb and Turner 1990, p458-464). This
particular model will also enable us to discuss one of the
`creation from nothing' claims made for mini-superspace quantum
cosmology.

We begin by selecting the 3-manifold to be $S^3$, and we only
consider metric tensors of the form

$$
ds^2 = R^2(d\chi^2 + \sin^2 \chi(d\theta^2 + \sin^2 \theta
d\phi^2)) \,.
$$

Each metric tensor of this type equips $S^3$ with a homogeneous
and isotropic Riemannian geometry. The scale factor $R \in
[0,\infty)$ is the only permitted degree of freedom in the spatial
geometry. In this particular model, it is also the only degree of
freedom, geometrical or non-geometrical. The matter field is
chosen to be a massive scalar field, fixed at some constant value
$\phi$; the value of the field is the same at each point of $S^3$.
The selection of a particular massive scalar field includes the
selection of a potential energy function $V(\phi)$. Hence, by
fixing a particular value $\phi$, one fixes a particular energy
density $\rho_\phi$. It will be helpful in what follows to define
a cosmological constant $\Lambda = 8\pi G\rho_\phi$ from the
energy density of the scalar field.

With the mini-superspace now defined, it is clear that a
wave-function will simply be a function $\Psi(R)$ of the possible
values for the scale factor. In general terms, the Wheeler-DeWitt
equation will have the form $(\nabla^2 - U(R)) \Psi(R) = 0$. With
a factor-ordering ambiguity $a$, the Wheeler-DeWitt operator has
the form:

$$
(R^{-a}\frac{\partial}{\partial R}R^a\frac{\partial}{\partial
R})-U(R) \,.
$$

With the potential $U(R)$ defined to be

$$
U(R) = \frac{9\pi^2}{4G^2}(R^2-\frac{\Lambda}{3}R^4) \,,
$$ and with $a$ set to $a = 0$, the Wheeler-DeWitt equation takes the
form, (Kolb and Turner, p459):

$$
\left [\frac{\partial^2}{\partial R^2}-
\frac{9\pi^2}{4G^2}(R^2-\frac{\Lambda}{3}R^4) \right ]\Psi(R)= 0
\,.
$$

This equation clearly resembles the time-independent Schrödinger
equation $H\Psi = E\Psi$ for a system constrained to move in
$[0,\infty)$, with a fixed total energy $E = 0$, and subject to
the potential $U(R)$.

This mini-superspace model has a profound relationship with de
Sitter space-time, a solution of the classical equations. To see
this, one introduces $R_0 = (\Lambda/3)^{-1/2} = (8\pi
G\rho_\phi/3)^{-1/2}$. One can then split the configuration space
$[0,\infty)$ into $0 < R \leq R_0$ and $R \geq R_0$. The potential
$U(R)$ is positive in the region $0 < R < R_0$, hence with the
total energy fixed at $E = 0$, this region is classically
forbidden. $R_0$ is the classical turning point, at which the
potential is zero. Hence, at $R = R_0$, the kinetic energy of a
classical system would have to be zero. In the region $R > R_0$,
the potential is negative, so $R \geq R_0$ is a classically
permitted region of the configuration space for the $E = 0$
system. Intriguingly, the potential $U(R)$ is zero at $R = 0$,
hence $R = 0$ is also a classically permitted configuration. A
classical system at $R = 0$ would have zero kinetic energy and
zero potential energy, and would remain at $R = 0$.

To understand the link between this mini-superspace model and de
Sitter space-time, recall that de Sitter space-time is
$\mathbb{R}^1 \times S^3$ equipped with the metric

$$
ds^2 = -dt^2 +R^2(t)(d\Omega_3^2)\,,
$$ where $R(t) = R_0 \cosh (R_0^{-1} t)$, and $d\Omega_3^2$ is the standard
metric on the 3-sphere.

De Sitter space-time can be treated as a solution of the Einstein
Field equations with a cosmological constant $\Lambda = 8\pi
G\rho_{vac}$. The vacuum energy density $\rho_{vac}$ corresponds
to the energy density $\rho_{\phi}$ of the scalar field in the
mini-superspace model.

If one foliates de Sitter space-time by the one-parameter family
of homogeneous t = constant spacelike hypersurfaces, then the
resulting family of spatial configurations corresponds to a curve
in the one-dimensional configuration space $[0,\infty)$ under
consideration. One has a classical universe which contracts from
the infinite past to a minimum scale factor of $R_0 =
(\Lambda/3)^{-1/2}$, and then expands without limit into the
infinite future. Thus, the region $[0,R_0)$ of the configuration
space is not entered by the classical de Sitter space-time. In the
mini-superspace model, this corresponds to the fact that $0 < R <
R_0$ is a classically forbidden region. The classical turning
point $R_0$ of the mini-superspace model corresponds to the
minimum radius of de Sitter space-time. The scale factor of de
Sitter space-time only occupies the classically permitted region
$R \geq R_0$ of the configuration space.

The transition to quantum theory involves the serious
consideration of all kinematically possible paths through a
configuration space. De Sitter space-time provides a path in
configuration space which is dynamically possible according to the
classical theory. By considering all kinematically possible paths,
the classically forbidden region $0 < R < R_0$ becomes
traversable. There are kinematically possible paths which do enter
$0 < R < R_0$. The most startling consequence of this is that a
quantum system which begins at $R = 0$, can tunnel through the
potential barrier, and reach $R > R_0$. Some quantum cosmologists
interpreted this as a prototypical model for the creation of the
universe \textit{ex nihilo}.

To actually calculate the probability of a system tunnelling from
$R = 0$ to $R > 0$, it is of course necessary to provide a
solution $\Psi(R)$ for the Wheeler-DeWitt equation of this
one-dimensional mini-superspace model. Vilenkin, Linde and
Hartle-Hawking make competing proposals for this wave-function,
(Vilenkin 1998).

For the $R > R_0$ region, the WKB solutions of the Wheeler-DeWitt
equation are

$$
\Psi_\pm(R > R_0) \propto \frac{1}{\sqrt{k(R)}}\exp \left (\pm i
\int^R_{R_0}k(R')dR'\mp i\frac{\pi}{4} \right ) \,,
$$ where $k(R) = \sqrt{(-U(R))}$ for the $E = 0$ mini-superspace model
under consideration.

For the $R < R_0$ region, the WKB solutions are

$$
\Psi_\pm(R < R_0) \propto \frac{1}{\sqrt{|k(R)|}}\exp \left (\pm
\int_R^{R_0}|k(R')| dR' \right ) \,.
$$

Vilenkin claims that the `ingoing' wave $\Psi_+(R > R_0)$
corresponds to a contracting universe, and that it is the
`outgoing' wave $\Psi_-(R > R_0)$, satisfying the condition
$i\Psi^{-1}\partial \Psi/\partial R > 0$, which corresponds to an
expanding universe. He claims that the wave-function should be
$\Psi_-(R > R_0)$ in the classically permitted region, and a
combination of ingoing and outgoing modes in the classically
forbidden region. From this, he calculates that the probability
for tunnelling through the potential barrier from $R = 0$ should
be $\sim exp(-|A_E|)$. Vilenkin, however, also makes the dubious
assertion that this provides the probability of creating an
expanding universe from `nothing'. Vilenkin equates $R = 0$ with
nothing in this context.

Linde proposes that the wave-function in the classically allowed
region should be a combination of incoming and outgoing modes,
$\frac{1}{2}[\Psi_+(R > R_0) + \Psi_-(R > R_0)]$, and should be
$\Psi_+(R < R_0)$ in the classically forbidden region. The
Hartle-Hawking proposal is also that the wave-function in the
classically allowed region should be a combination of incoming and
outgoing modes, $\Psi_+(R > R_0) - \Psi_-(R > R_0)$, and should be
$\Psi_-(R < R_0)$ in the classically forbidden region.
Hartle-Hawking calculate the probability of a transition from $R =
0$ to $R = R_0$ by means of a signature change scenario. They take
the Riemannian four-sphere $S^4$, and they remove one hemisphere,
joining the equator to Lorentzian de Sitter space-time at its
minimum radius $R_0$. They take the Euclidean action $A_E$ of the
compact Riemannian region, and they assert that the probability of
a transition from $R = 0$ to $R = R_0$ is $\sim exp(-A_E)$.

The four-sphere region, as a Riemannian solution of the classical
vacuum field equations,

$$
R_{\mu \nu} = \Lambda g_{\mu \nu} \,,
$$ a so-called `gravitational instanton', is
automatically geodesically complete. As pointed out in Section 3,
this is a generic outcome of the Euclidean approach, and motivates
the suggestion that an initial singularity can be avoided in
Euclidean quantum cosmology. Given, however, that mini-superspace
quantum cosmology replaces an individual space-time manifold with
a wave-function, the issue of a singularity in the geometry has
been by-passed; a geometric object has been replaced by an object
from a function space. The only sense in which the question of
geometric singularities might re-emerge is with respect to the
classical paths in configuration space which can, under certain
conditions, be derived from the wave-function.

The approaches of Vilenkin, Linde and Hartle-Hawking are each
vulnerable to the following objection: In quantum mechanics, there
is a probabilistic propensity for a system to make a transition
from one side of a potential barrier to the other. If a transition
takes place, the common interpretation is that state reduction has
taken place, and the probabilistic propensities are replaced by a
definite position on the other side of the potential barrier.
Given that the universe is a closed system, the concept of state
vector reduction is not obviously applicable in quantum cosmology.
Hence, unless one subscribes to an interpretation of quantum
theory which rejects state vector reduction, the concept of
tunnelling through a potential barrier cannot explain the creation
of the universe. In general, therefore, quantum cosmologists have
been forced to explore and apply various `no-collapse'
interpretations of quantum theory. It may be, of course, that a
no-collapse interpretation of quantum theory is the correct one to
take, but it is important to emphasise the dependence of quantum
cosmology explanations upon non-standard interpretations of
quantum theory.

Vilenkin's attempt to equate `nothing' with $R = 0$ also seems
completely incorrect. If one considers $S^3$ with a metric

$$
ds^2 = R^2(d\chi^2 + \sin^2 \chi(d\theta^2 + \sin^2 \theta
d\phi^2))\,,
$$ then setting $R = 0$ simply removes the geometry from $S^3$. One
still has the 3-manifold $S^3$, a 3-dimensional space without any
geometry. Thus, Vilenkin's theory might conceivably show that the
universe was created from a 3-dimensional space, bereft of
geometry and matter, but it cannot show that the universe was
created from nothing. Moreover, in this one-dimensional
mini-superspace model, the scalar field is fixed at a constant
value $\phi$ on $S^3$. Hence, when $R = 0$, the scalar field still
presumably exists on $S^3$. One has a 3-manifold $S^3$, equipped
with a matter field; this is far from being nothing.

Prugovecki associates Vilenkin's scenario with Tryon's idea for
the creation of the universe as a fluctuation of `nothing'.
Prugovecki argues that ``the concept of a wave function,
representing a quantum particle, `tunneling through' the potential
barrier to which another system of existing quantum particles
gives rise, is operationally well-defined, and it makes physical
sense; however, what is the possible physical meaning of Nothing
tunneling through a potential barrier produced by Nothing, in
order to `create' our Universe?" (Prugovecki 1992, p454).

Prugovecki (1992, p481, note 33) asks, with justification, ``what
is it that is supposedly `tunneling', and through a barrier of
what does that purported `tunneling' take place in the
Tryon-Vilenkin `scenario'?" .

\subsection{Vilenkin's tunnelling boundary condition}

Vilenkin's stipulation that $i\Psi^{-1}\partial \Psi/\partial R >
0$ for the one-dimensional mini-superspace model discussed above,
is a special case of his `tunnelling boundary condition' on the
wave-function of the universe. To understand Vilenkin's boundary
condition in greater generality, we shall now review, clarify, and
critically analyse a mini-superspace model with two degrees of
freedom. This mini-superspace model is of particular interest
because it has been used to address the question of whether
quantum cosmology can predict the false vacuum necessary for
inflation to take place. The notion that Vilenkin's tunnelling
boundary condition specifies the probability of creating a
universe from nothing, will continue to receive attention, and in
particular, Vilenkin's notion of a boundary will be subjected to
critical scrutiny.

We again select the 3-manifold $\Sigma$ to be $S^3$, and we again
consider only the metrics on $S^3$ of the form

$$
ds^2 = R^2(d\chi^2 + \sin^2 \chi(d\theta^2 + \sin^2 \theta
d\phi^2)) \,.
$$

The scale factor $R \in [0,\infty)$ is therefore the only
contemplated degree of freedom in the spatial geometry.

We select the matter field on $S^3$ to be a massive scalar field
$\phi$ of constant value across $\Sigma$. The value of $\phi$ is
the degree of freedom in the matter field. The potential energy
density $V(\phi)$ is considered to be a function of $\phi$.
Different forms of the function $V(\phi)$ yield different
mini-superspace models. The spatial geometry is fixed to be
homogeneous and isotropic, and the matter field is also clearly
homogeneous.

A wave-function in this model therefore has the form
$\Psi(R,\phi)$, and the Wheeler-DeWitt equation will be, (Kolb and
Turner, p463):

$$
\left [ R^{-a}\frac{\partial}{\partial
R}R^a\frac{\partial}{\partial R}-\frac{1}{R^2}\frac{3}{4\pi
G}\frac{\partial^2}{\partial \phi^2}-U(R,\phi) \right ]\Psi(R,
\phi)= 0 \,,
$$ where $a$ is a factor ordering ambiguity, and where the
`superpotential' $U(R,\phi)$ is given by

$$
U(R,\phi) = \frac{9\pi^2}{4G^2}\left (R^2-R^4\frac{8\pi
G}{3}V(\phi)\right ) \,.
$$

The Wheeler-DeWitt equation here clearly has the form $(\nabla^2 -
U)\Psi = 0$, where $\nabla^2$ is a Laplacian. Once again the
Wheeler-DeWitt equation resembles the time-independent Schrödinger
equation $H\Psi = E\Psi$ for a particle of fixed total energy $E =
0$, moving in a potential $U$.

Vilenkin's `tunnelling' boundary condition, if it could be shown
to be meaningful, would be a genuine boundary condition on the
wave-function $\Psi$. Vilenkin conceives that superspace has a
boundary, and he attempts to identify a unique wave-function
$\Psi$ by its behaviour on the boundary of superspace. Thus, the
boundary of Vilenkin's boundary condition is a topological
boundary of the domain of the wave-function. Attempts to
practically implement Vilenkin's boundary condition have been
restricted to mini-superspace models, where the boundary is the
boundary of the mini-superspace.

Vilenkin's notion of the boundary of superspace is rather unclear.
He asserts that ``the boundary of superspace can be thought of as
consisting of singular configurations which have some points or
regions with infinite three-curvature or with infinite $\phi$ or
$(\partial_i\phi)^2$, as well as configurations of infinite
three-volume," (Vilenkin 1988, p889).\footnote{Vilenkin only
contemplates compact 3-topology, so infinite volume is considered
to be a kind of pathology.} This is a wholly inadequate definition
of the boundary of superspace. Suppose that we have fixed a
compact, orientable 3-manifold $\Sigma$, and that we introduce the
configuration space $\mathscr{C}(\Sigma)$ of all Riemannian metric
tensor fields on $\Sigma$. Let $T^*\Sigma$ denote the cotangent
bundle of $\Sigma$, and let $\odot^2T^*\Sigma$ denote the 2-fold
symmetric tensor product of the cotangent bundle.
$\mathscr{C}(\Sigma)$ is an open positive cone in the set of
smooth cross sections $C^\infty(\odot^2T^*\Sigma)$ of the vector
bundle $\odot^2T^*\Sigma$. One can now ask: Should the boundary of
the configuration space be the topological boundary of
$\mathscr{C}(\Sigma)$, considered as an open subset in the
topological space $C^\infty(\odot^2T^*\Sigma)$? Should one then
take the quotient of this boundary with respect to the
$Diff(\Sigma)$ action, to find the boundary of the superspace
$\mathscr{C}(\Sigma)/Diff(\Sigma)$?

The boundary of an open subset $\mathcal{X}$ of a topological
space $\mathcal{T}$, is the set of points in the closure of
$\mathcal{X}$ which do not belong to $\mathcal{X}$. The closure of
an open subset $\mathcal{X}$ is the union of $\mathcal{X}$ with
the set of accumulation points of $\mathcal{X}$ which do not
belong to $\mathcal{X}$. Thus, the boundary of an open subset
$\mathcal{X}$ is the set of those accumulation points which do not
belong to $\mathcal{X}$. An accumulation point $x$ of a subset
$\mathcal{X}$ is a point for which every neighbourhood contains
points of $\mathcal{X}$ other than $x$. For first-countable
topological spaces, of which manifolds are particular cases, a
point $x$ is an accumulation point of a subset $\mathcal{X}$ if
and only if there is a sequence of points in $\mathcal{X}-x$ which
converges to $x$.

The boundary of $\mathscr{C}(\Sigma)$ will be dependent upon the
topological space that $\mathscr{C}(\Sigma)$ is considered to be
an open subset of. Even if one fixes the set $\mathcal{T}$ that
$\mathscr{C}(\Sigma)$ is considered to be a subset of, one can
vary the topology of $\mathcal{T}$. The closure of
$\mathscr{C}(\Sigma)$, and therefore its boundary, will be
different in different topologies. In general, the coarser the
topology of the set in which $\mathscr{C}(\Sigma)$ is considered
to be a subspace, the larger the closure will be. Vilenkin
discusses none of these questions. Moreover, no matter what the
topology of $\mathcal{T}$, the accumulation points of
$\mathscr{C}(\Sigma)$ in $\mathcal{T}$ will not correspond to
tensor fields which diverge at points or regions of $\Sigma$.
Infinity, $\infty$, is not an element in the field of real
numbers, and we are dealing with a module
$C^\infty(\odot^2T^*\Sigma)$ of tensor fields over the ring of
real-valued scalar fields on the manifold $\Sigma$.

However, to understand Vilenkin's notion of a superspace boundary,
one might be able to densely embed $\mathscr{C}(\Sigma)$ in an
appropriate space constructed from sequences of points in
$\mathscr{C}(\Sigma)$. Each $\gamma \in \mathscr{C}(\Sigma)$ could
be mapped to an equivalence class of sequences which converge to
$\gamma$. Sequences which tend towards infinite volume, infinite
curvature, or infinite matter field values, for example, do not
converge to points of $\mathscr{C}(\Sigma)$, but might form the
boundary of $\mathscr{C}(\Sigma)$ in a suitable dense embedding.

For example, take an arbitrary point $\gamma \in
\mathscr{C}(\Sigma)$ in the configuration space, and consider the
sequence of points $S = \{P(N) = N^2\gamma : N \in
\mathbb{Z}_+\}$. $\mathbb{Z}_+$ is the set of positive integers.
The sequence has no limiting point in $\mathscr{C}(\Sigma)$ as $N
\rightarrow \infty$. It is a sequence of homothetic 3-geometries
in which the volume grows without limit. For any integer $M$
between $1$ and $\infty$, one can get a subsequence of $S$ by
taking only the members of $S$ up to $P(M)$. Each such sequence
$S_M$ does converge to a point of $\mathscr{C}(\Sigma)$; it
converges to $P(M)$. The sequence of $S_M$-sequences converges to
the sequence S. That is, $\lim_{M \rightarrow \infty} S_M = S$.
Thus, by this construction, one can treat $S$ as a boundary point
of $\mathscr{C}(\Sigma)$.

When Vilenkin speaks of a configuration of infinite volume, one
might interpret this to be a metaphorical way of referring to a
sequence of configurations in which the volume increases without
limit. It could only be metaphorical because the volume is finite
in each member of the sequence. One might treat Vilenkin's talk of
infinite curvature and infinite matter field values similarly.

Vilenkin attempts to divide the boundary of superspace into a
`singular' boundary and a `regular' boundary. He defines every
regular boundary configuration to be a `critical' slice of a
topology-changing 4-geometry. The slicing, or `foliation', is
defined here by a Morse function $f(x)$, and a critical slice
contains critical points of the Morse function, points $x_0$ at
which $\partial_\mu f(x_0) = 0$, (Vilenkin 1994, p2588-2589). The
3-geometry on such a slice is degenerate at the isolated critical
points. Vilenkin defines the singular boundary to be composed of
those configurations with infinite volume, curvature etc, which
cannot be embedded as a critical slice in a 4-geometry.

To express his boundary condition, Vilenkin employs a probability
current density

$$
J = i/2(\Psi^*\nabla \Psi - \Psi \nabla \Psi^*) \,,
$$ and he states that ``this current can be identified with the
probability flux in superspace," (Vilenkin 1988, p889). The
tunnelling boundary condition is that the wave-function of the
universe should only include `outgoing modes' at the singular
boundary, ``carrying flux out of superspace," (Vilenkin 1988,
p889). Vilenkin is proposing that the probability flux vector
field $J$, associated with $\Psi$, must point out of superspace at
the singular boundary. According to Vilenkin, this corresponds to
a non-singular beginning to the universe.

Vilenkin states that a WKB wave-function can be written as a
superposition

$$
\Psi = \sum_n C_n e^{iS_n} \,,
$$ where the $S_n$ are rapidly varying functions, each of which
satisfies the time-independent Hamilton-Jacobi equation on
mini-superspace:

$$
\| (\nabla S_n)\|^2 + U = 0 \,.
$$ Vilenkin asserts that the current for the nth term is

$$
J_n = - | C_n | ^2 \nabla S_n \,,
$$ and that the tunnelling boundary condition requires that the
vector fields $-\nabla S_n$ should point out of superspace at the
singular boundary, (Vilenkin 1988, p890). He states that each
function $S_n$ defines a congruence of `classical trajectories' in
the mini-superspace, the integral curves of the vector fields
$-\nabla S_n$. The tunnelling boundary condition means that these
classical paths can end at the singular boundary, but not begin
there.

Vilenkin recognizes that one need not restrict attention to the
(Riemannian) geometries on a fixed 3-manifold: ``We can define the
extended superspace$\ldots$including all possible topologies. It
can be split into topological sectors, with all metrics in the
same sector having the same topology," (1994, p2588). If one
restricts attention to 3-geometries which are non-degenerate at
all points, then the superspace of all possible 3-geometries is a
disconnected topological space, with each component corresponding
to geometries having the same topology. Vilenkin, in contrast,
seems to envisage a connected space, with the metrics of different
topology occupying disjoint open subsets, separated from each
other by boundaries. In line with this, he originally held that
``topology changing transitions$\ldots$occur through the boundary
of the corresponding superspace sectors," (1994, p2588). The
boundaries between different topological sectors are Vilenkin's
regular boundaries, each point of which is the critical slice of a
topology-changing 4-geometry, (1994, p2589). One can foliate a
topology-changing space-time by a one-parameter family of
spacelike hypersurfaces if one permits the metric on some of those
spacelike hypersurfaces to be degenerate at isolated points (Borde
1994). If one enlarges the space of 3-geometries to include those
which are degenerate at isolated points, then one can represent a
topology-changing space-time as a curve in this enlarged
configuration space.

Whilst the outgoing-wave boundary condition was imposed on the
singular part of the boundary, ``the boundary condition on [the]
regular boundary was supposed to enforce conservation of
probability flux as it flows from one topological sector to
another," (Vilenkin 1994, p2589). Vilenkin held that there is a
boundary between the `null topological sector', which contains
nothing, and a sector such as that containing all the
$S^3$-geometries, (1994, p2588). He believed this boundary to be
part of the regular boundary: ``the probability flux is injected
into superspace through the boundary with the null sector; it then
flows between different topological sectors through the regular
boundaries, and finally flows out of superspace through the
singular boundary," (1994, p2589).

Vilenkin later held that ``topology change does not necessarily
occur between configurations at the boundaries of superspace
sectors, but generally involves configurations in the interiors of
these sectors," (1994, p2589). Whilst topology change must occur
through a critical slice, Vilenkin contends that such critical
slices can lie in the interior of superspace sectors. It seems,
then, that every regular boundary configuration is a critical
slice, but not every critical slice is part of the regular
boundary.

In terms of an $S^3$ mini-superspace model with a scale factor $a$
and matter field $\phi$, each point in the mini-superspace is a
pair $(a,\phi)$. In terms of an $S^3$ mini-superspace model with a
scale factor $a$ mapped to $\alpha = \ln a$, and a matter field
$\phi$, each point in this mini-superspace is a pair
$(\alpha,\phi)$. The mini-superspace of pairs $(a,\phi)$ is the
manifold $(0,\infty) \times (-\infty,+\infty)$, whilst the
mini-superspace of pairs $(\alpha,\phi)$ is the manifold
$(-\infty,+\infty) \times (-\infty,+\infty)$. Vilenkin asserts
(1994, p2588) that ``the surface $\alpha = -\infty$, $| \phi | <
\infty$ can be thought of" as the boundary between the ``null
topological sector" and the sector associated with $S^3$. Thus,
using the interpretation of Vilenkin's boundary-concept suggested
above, each sequence of points $(\alpha_n,\phi_n)$ in which
$\lim_{n \rightarrow \infty} \alpha_n = -\infty$, is a point of
the (regular) boundary with the null sector. In other words, a
sequence of configurations in which the scale factor $a =
\exp(\alpha)$ tends to zero, and the matter field $\phi$ converges
to a finite value, is a point of the (regular) boundary with the
null sector. Sequences in which $\lim_{n \rightarrow \infty}
\alpha_n = -\infty$ \textbf{and} $\lim_{n \rightarrow \infty} |
\phi_n | = \infty$, or sequences in which $\lim_{n \rightarrow
\infty} \alpha_n = +\infty$, correspond to points in the singular
boundary.\footnote{Sequences in which $\lim_{n \rightarrow \infty}
\alpha_n = +\infty$, are sequences in which the scale factor, and
thus the volume of space, increase without limit. Sequences in
which $\lim_{n \rightarrow \infty} | \phi_n | = \infty$, are
sequences in which the matter field either increases without
limit, or decreases without limit. It is worth emphasizing once
again that each point in any of these sequences would be a point
of the mini-superspace, hence the scale factor and matter field
would be finite for each point in such a sequence.}

Vilenkin asserts that the probability flux is ``injected" through
the regular boundary $\alpha = -\infty$, $| \phi | < \infty$, the
boundary with the `null topological sector', and ``flows out of
superspace through the remaining boundary ($\alpha \rightarrow
-\infty$ with $| \phi | \rightarrow \infty$, or $\alpha
\rightarrow +\infty)$," (ibid.).

However, the `null topological sector' is just the empty set, and
there is no reason to think of it as sharing a boundary with a
non-empty set of geometries. Hence, the integral curves on the
configuration space (or superspace) which result from Vilenkin's
boundary condition on the wave function, should not be interpreted
as describing an ensemble of universes which are created from
nothing.

Vilenkin's ideas are intriguing, but it is not established that
they can be meaningfully extended to infinite-dimensional
superspaces, his notion of the boundary of a superspace is
unsatisfactory, and a wave-function satisfying the tunnelling
boundary condition cannot be interpreted as describing creation
from nothing.

\end{document}